\documentclass[twocolumn]{pasj00}
\SetRunningHead{A. Yoshino and T. Ichikawa}{Colors and $M/L$ ratios of
bulges and disks}

\title{Colors and Mass-to-Light Ratios of Bulges and Disks of Nearby
Spiral Galaxies}
\author{Akira \textsc{Yoshino}}
\affil{%
   National Astronomical Observatory of Japan, \\
   Osawa 2-21-1, Mitaka, Tokyo 181--8588}
\email{yoshino.akira@nao.ac.jp}

\and 
\author{Takashi \textsc{Ichikawa}}

\affil{%
   Astronomical Institute, Tohoku University, \\
   Aoba, Sendai, Miyagi 980--8578}
\email{ichikawa@astr.tohoku.ac.jp}

\Received{2004/10/20}
\Accepted{2008/02/15}
\Published{$\langle$publication date$\rangle$}
\begin{document}
\maketitle

\begin{abstract}
We investigate colors and mass-to-light ratios ($M/L$s) of the bulges
 and disks for 28 nearby spiral galaxies with various morphological
 types of Sab to Scd, using images in optical and near-infrared ($V$,
 $I$, and $J$) bands and published rotation curves. It is shown that
 the observed colors and $M/L$s generally agree with the galaxy
 formation model with an exponentially declining star formation rate
 and shallow slope (ex. Scalo) initial mass function (IMF) for both
 the bulges and the disks. We find that the bulge $M/L$ is generally
 higher than the disk $M/L$ and that the galaxies with larger
 bulge-to-total luminosity ratio tend to have a smaller bulge
 $M/L$. The fact indicates that the luminosity-weighted average age of
 bulges for early-type spirals is younger than that of later-type
 spirals. These results support a formation scenario that produces 
 young stars for the bulges of middle-type and early-type spirals.
\end{abstract}

\section{Introduction}

The origin of bulges is a clue to understanding galaxy formation and
evolution.  Previously two main scenarios: monolithic collapse and
secular formation models for the bulge formation have been proposed.
In monolithic collapse scenarios, bulges were formed rapidly from
primordial gas as well as elliptical galaxies and then disks were
grown gradually (e.g., \cite{eggen1962}; \cite{arimoto1987}). On the
other hand, in secular formation scenarios, bulges were formed by
redistribution or secular heating of disk stars (\cite{zhang1999}),
bar dissolution (\cite{kormendy1979}; \cite{norman1996}), or satellite
galaxies falling into the host galaxy center region
(\cite{aguerri2001}).  Thus in these scenarios bulges were formed
gradually in disks and consequently do not have uniform age; the
average age of some bulges would be as young as that of disks. The
bulges formed from secular process are called ``pseudo bulges'' (e.g.,
\cite{kormendy2004}, \cite{laurikainen2007}) in contrast to
``classical bulges'', which formed rapidly in high-z.  Furthermore,
\citet{kauffmann1996} predicted that the bulges of early-type spirals
is younger than the bulges of late-type spirals from her semi-analytic
merger models of galaxy formation in a flat cold dark matter universe.

The colors and mass-to-light ratios ($M/L$s) of bulge and disk are
fundamental properties to investigate the origin. In general, both the
color index and the $M/L$ are correlated with the age. However, they
are also affected by the metallicity (age-metallicity degeneracy)
(e.g., \cite{worthey1994}; \cite{kodama1997}). Moreover, the dust
absorption also causes seemingly higher $M/L$ and redder
color. Therefore the observations for colors and $M/L$s in a wide
wavelength range are needed to resolve the degeneracy.

Several galaxy formation models predicting the colors and $M/L$s have
been reported. Worthey (1994, hereafter W94) constructed the models
with various single starburst-age and the initial mass function (IMF)
of \citet{salpeter1955}. He showed that the $M/L$ in $I_c$ band
(hereafter we call it simply $I$ band) depends strongly on the age of
the system but is nearly independent of the metallicity. Hence he has
recommended the use of $M/L$ in the $I$ band to estimate the age of
the stellar system.

\citet{bell2001} (hereafter BD01) and recently \citet{portinari2004}
presented the galaxy evolution models with various e-folding
time-scale of star formation rate (SFR) and various IMFs. In the
context of their schemes, a shallow slope IMF or a large secondary
starburst (e.g., 10\% fraction of the total stellar mass) makes $M/L$
smaller than that from single starburst at high-z and steep slope IMF,
because the shallow slope IMF reduces the number of low-mass stars,
and the large secondary starburst produces the young stars. Comparing
the models with the observed disk $M/L$s of the spiral galaxies in
Ursa Major cluster of galaxies, BD01 pointed out that the observed
disk $M/L$s are generally consistent with the model with shallow slope
IMF. However, the bulge $M/L$s were not explicitly investigated in
BD01.

In order to investigate bulge properties, the spiral galaxies should
be decomposed into bulge and disk components. Moreover, to obtain
bulge $M/L$s, accurate rotation curves around the galactic center
region are necessary. Previously many authors (e.g., \cite{kent1986};
\yearcite{kent1987}; \yearcite{kent1988}; Moriondo et al. 1998b)
attempted to obtain the colors and $M/L$s of bulges and disks, and
investigated the correlation between $M/L$s and the dark matter mass,
and morphological type of galaxies.  The rotation curves used in
\authorcite{kent1986} (\yearcite{kent1986}, \yearcite{kent1987},
\yearcite{kent1988}) and \citet{moriondo1998b} were obtained with
H\emissiontype{I} observations (e.g., \cite{carignan1985};
\cite{bosma1981}) or H${\alpha}$ observations of \citet{rubin1985}.
However, the bulge $M/L$s may have been underestimated (e.g.,
\cite{kent1988}; Moriondo et al. 1998b), because the rotation curves
used in their analyses may not measure the true circular velocity in
the central region of the galaxies. \citet{sofue1996} claimed that the
H${\alpha}$ or other ionized gas rotation curves are often influenced
by non-circular velocity component (e.g., inflow, outflow or streaming
motions) rather than an ordered circular rotation, thus the rotation
curve in inner part ($\le$1 kpc) tends to have the slowly rising
feature (e.g., \cite{fillmore1986}).  Also, since the
H\emissiontype{I} gas cannot be well detected in the region of the
central a few kilopersecs, the rotation velocity was generally
smoothed around the central region by observers, the resulting
rotation curves have an apparently solid-body (slowly rising) feature
(\cite{sofue1996}).  Since the $M/L$ is proportional to the square of
rotation velocity, the $M/L$ in bulge region is estimated
significantly low using the slowly rising rotation curve.

On the other hand, the observation of CO line-emission gives more
accurate velocity field at the central region of galaxies because CO
molecular gas, which is one of cold interstellar medium, concentrates
in the central a few kpc region and is easily detected in the region
rather than H\emissiontype{I}, H${\alpha}$ or other ionized
gas (Sofue 1996).  Therefore the CO rotation curves should represent
more accurately the circular rotation velocity and are more desirable
to obtain bulge $M/L$s than other rotation curves.  In this paper we
use the CO rotation curves of \citet{sofue1997} and partly
\citet{sofue2003}.  They observed the CO rotation curves around the
centers of nearby galaxies with 45 m telescope at Nobeyama Radio
Observatory (NRO) with a resolution of 15 arcsec in \citet{sofue1997}
and Nobeyama Millimeter-wave Array (NMA) with 2-4 arcsec in
\citet{sofue2003}, and connected to H\emissiontype{I} rotation curves
for outer regions of galaxies.  The resolution of 15 or 2-4 arcsec is
not high but enough to discuss bulge properties especially for large
nearby galaxies. Indeed, most of the CO rotation curves in
\citet{sofue1997} and \citet{sofue2003} have revealed steeply rising
features, which may be originated from the mass concentration of bulge
region, thus the bulge $M/L$ would be higher than previously estimated. 

Our motivation in this paper is to evaluate accurately colors and
$M/L$s of bulges and disks and to study the ages. First, galaxy images
are decomposed 2-dimensionally into bulge and disk models. $M/L$s for
bulge and disk are obtained by the fitting of rotation velocity models
to the observational ones. In the rotation curve fitting, we adopt
``maximum bulge plus disk solution", which assumes that the inner
rotation curve is originated only by the luminous matter. The method
is based on \citet{albada1985} and \citet{kent1986}. Finally, these
results are compared with the stellar population models of W94 and
BD01.

In next section, we describe the data and the reductions. The
two-dimensional decomposition method for the galaxy images and the
fitting method of the velocity model to the rotation velocity data are
explained in detail in section 3. We carefully examine the effect of
noise and point spread function (PSF) for the images and the aperture
effect on the rotation curve measurement on the basis of Monte Carlo
simulations with artificial galaxies. In section 4 we show the results
for the colors and $M/L$s of bulge and disk components of the sample
galaxies. The results are discussed in section 5 in the context of age
and formation of the components. Finally we conclude in section 6.

\section{Data and Reductions}

\subsection{The Sample Galaxies}

The sample galaxies are listed in table 1. They are basically taken
from \citet{sofue1997} and \citet{sofue2003}. The galaxies in
\citet{sofue1997} were chosen that (a) the angular size is large
enough in order to obtain a sufficiently high linear scale resolution
in the central regions; (b) the disk is mildly tilted in order for an
accurate correction for the inclination to derive the rotation
velocity; (c) no high sensitivity data have been obtained yet with the
45 m telescope or the NMA; and (d) the CO-line emission is
sufficiently strong or IRAS 60 and 100 $\mu$m fluxes are higher than
several Jy (Sofue 1996).  These rotation curves are observed using NRO
45 m with 15 arcsec resolution.  Some rotation curves (NGC 4192, NGC
4535, NGC 4536, NGC 4548, and NGC 4569) are taken from
\citet{sofue2003}, which targets Virgo cluster galaxies using NMA with
2-4 arcsec resolution.  The accuracy of velocity is about
$\rm{15-20~km~s^{-1}}$ for both data of \citet{sofue1997} and
\citet{sofue2003}.  We searched for the wide-field imaging data
corresponding to the rotation curve data as much as possible.  As the
result, we obtained the three bands ($V$, $I$, and $J$) imaging data
from SMOKA\footnote{Based on data collected at Kiso observatory
(University of Tokyo) and obtained from data archive at Astronomy Data
Center, which are operated by the National Astronomical Observatory of
Japan.} archive system. The numbers of sample are 23 for $V$ band, 24
for $I$ bands, and 20 for $J$ band. The $V$ and $I$ band images were
observed with 105 cm Schmidt telescope with a 1k$\times$1k CCD camera
at the Kiso Observatory, University of Tokyo except NGC 3198, NGC
3521, NGC 4258, NGC 4321, NGC 4535, NGC 4536, NGC 4548, NGC 5457, and
NGC 7331, which were observed with a 2k$\times$2k CCD camera. The $J$
band images for 20 galaxies were observed with KONIC
(\cite{itoh1995}). The full width half maximum (FWHM) of seeing is
$3.8\pm0.9$ arcsec for all bands. The exposure time is typically 300s
for 1k CCD, 1200s for 2k CCD, and 1800s for KONIC. The information of
the telescope and the cameras are listed in table 2.

The data of morphology, isophotal diameter at 25.0 mag arcsec$^{-2}$ in
$B$ band ($D_{25}$) and total magnitude in $B$ band ($B_T$) in table 1
are taken from ``The third reference catalogue of bright galaxies''
(RC3) (\cite{devaucouleurs1991}). The distances for 13 galaxies in
table 1 are based on the observation of Cepheid, and those for the
remainder are taken from \citet{tully1988}.

\begin{table*}
\begin{center} 
\caption{\bf The list of the galaxies.} 
\begin{tabular}{ccccccc}    
\hline\hline
 Name & Type & Distance     & $D_{25}$ & $B_T$ & Band  & RC \\
  (1)  & (2) & (3)          & (4)      & (5)   & (6)   & (7) \\
\hline
 NGC 253 & SAB(s)c & 3.0\footnotemark[(1)] & 27.5& 8.04 & $V$,$I$,$J$ & 1 \\
 NGC 891 & SA(s)b: & 9.6\footnotemark[(1)] & 13.5& 10.81 & $V$,$I$,$J$ & 1 \\
 NGC 1068 & RSA(rs)b & 14.4\footnotemark[(1)]  & 7.1 & 9.61 & $V$,$I$ & 1 \\ 
 NGC 1808 & RSAB(s)a & 10.8\footnotemark[(1)] & 6.5 & 10.76 & $J$ & 1 \\
 NGC 2403 & SAB(s)cd &  3.2\footnotemark[(2)]  & 21.9 & 8.93  & $V$,$I$ & 1 \\
 NGC 2841 & SA(r)b: & 14.1\footnotemark[(3)]   &  8.1 & 10.09  & $V$,$I$,$J$ &  1 \\
 NGC 2903 &  SAB(rs)bc  &  6.3\footnotemark[(1)]  & 12.6  & 9.68  & $V$,$I$,$J$ & 1 \\
 NGC 3031 & SA(s)ab & 3.63\footnotemark[(4)]   & 26.9  & 7.89  &
 $V$,$I$,$J$ & 1 \\
 NGC 3079 & SB(s)c & 20.4\footnotemark[(1)] & 7.9 & 11.54  & $V$,$I$,$J$ & 1 \\
 NGC 3198 & SB(rs)c  &  14.5\footnotemark[(5)]  & 8.5 & 10.87  & $V$,$I$,$J$ & 1 \\
 NGC 3521 & SAB(rs)bc & 7.2\footnotemark[(1)]   & 11.0 & 9.83  & $V$,$I$,$J$ & 1  \\
 NGC 3628 & SAb pec & 7.7\footnotemark[(1)]  & 14.8  & 10.28  & $J$ & 1 \\ 
 NGC 4192 & SAB(s)ab & 16.1\footnotemark[(6)]   & 9.8 & 10.95  &
 $V$,$I$ & 2 \\
 NGC 4258 & SAB(s)bc &  7.8\footnotemark[(7)]  & 18.6 & 9.10  &
 $V$,$I$,$J$ & 1 \\
 NGC 4303 & SAB(rs)bc & 15.2\footnotemark[(1)]  & 6.5  & 10.18  & $J$ & 1 \\
 NGC 4321 & SAB(s)bc  & 16.1\footnotemark[(6)]   & 7.4 & 10.05  & $V$,$I$,$J$ & 1 \\
 NGC 4535 & SAB(s)c & 16.1\footnotemark[(6)]   & 7.1 & 10.59  &
 $V$,$I$,$J$ & 2 \\
 NGC 4536 & SAB(rs)bc & 16.1\footnotemark[(6)]   & 7.6 & 11.16 & $V$,$I$,$J$ & 2 \\
 NGC 4548 & SB(rs)b & 16.1\footnotemark[(6)]   & 5.4 & 10.96  &
 $V$,$I$ & 2 \\
 NGC 4565 & SA(s)b: & 9.7\footnotemark[(1)]   & 15.9 & 10.42  &
 $V$,$I$ & 1 \\
 NGC 4569 & SAB(rs)ab  & 16.1\footnotemark[(6)]  & 9.5 & 10.26  & $V$,$I$ & 2 \\
 NGC 4631 & SB(s)d & 6.9\footnotemark[(1)]  &  15.5  & 9.75  & $J$ & 1 \\
 NGC 4736 & RSA(r)ab  & 4.3\footnotemark[(1)]  & 11.2  & 8.99  & $I$,$J$ & 1 \\
 NGC 5194 & SA(s)bc pec &  7.7\footnotemark[(1)]  & 11.2 & 8.96  &
 $V$,$I$ & 1 \\
 NGC 5457 & SAB(rs)cd    & 7.4\footnotemark[(8)]   & 28.8  & 8.31  & $V$,$I$ & 1 \\  
 NGC 5907 & SA(s)c: & 14.9\footnotemark[(1)]   & 12.8  & 11.12  & $V$,$I$,$J$ & 1 \\
 NGC 6946 & SAB(rs)cd &  5.5\footnotemark[(1)]  & 11.5  & 9.61  & $V$,$I$,$J$  & 1 \\
 NGC 7331 & SA(s)b &  15.1\footnotemark[(9)]  & 10.5  & 10.35  & $V$,$I$,$J$ & 1 \\
\hline
 
\multicolumn{6}{@{}l@{}}{\hbox to 0pt{\parbox{180mm}{\footnotesize
 \par\noindent

Col.(1) : NGC Number.  Col.(2) : Type taken from RC3.

Col.(3) : Distance [Mpc].  \footnotemark[(1)] \citet{tully1988}
 \footnotemark[(2)] \citet{freedman1988}

\footnotemark[(3)] \citet{macri2001} \footnotemark[(4)]
 \citet{freedman1994} \footnotemark[(5)] \citet{kelson1999}

 \footnotemark[(6)] \citet{ferrarese1996} \footnotemark[(7)]
\citet{newman2001} \footnotemark[(8)] \citet{kelson1996}

\footnotemark[(9)] \citet{hughes1998}

Col.(4) : Diameter taken from RC3 [arcmin].  Col.(5) : Total magnitude
taken from RC3 [mag].  

Col.(6) : Band for imaging.  Col.(7) : Reference of the Rotation
Curve. 1 is \citet{sofue1997},

2 is \citet{sofue2003}.

}\hss}}
\end{tabular}
\end{center} 
 \label{tab.1}
\end{table*} 

\begin{table}[hpbt]
\begin{center}
\caption{Specifications of telescope and cameras.} 
\begin{tabular}{cccc}
\hline\hline 
Location  &      Kiso Observatory\\
Telescope &     105 cm Schmidt Telescope\\
Camera &  1kCCD (TI Japan TC215 1024$\times$1024)\\
         &   2kCCD (SITe 2048$\times$2048)\\
         &   KONIC (PtSi 1040$\times$1040)\\
Resolution &  0.75 arcsec/pixel (1kCCD)\\
           &   1.5 arcsec/pixel (2kCCD)\\
           &   1.06 arcsec/pixel (KONIC)\\
Field of view &  12.5 arcmin$\times$12.5 arcmin (1kCCD)\\
           &   50 arcmin$\times$50 arcmin (2kCCD)\\
           &   18 arcmin$\times$18 arcmin (KONIC)\\
\hline 
\end{tabular}
\end{center}
 \label{tab.2}
 \end{table}

\subsection{Reduction and Calibration for Images}

The reduction for image data follows standard procedure. We therefore
describe only briefly the method. Data analysis were in part carried
out on general common use computer system at the Astronomy Data
Center, ADC, of the National Astronomical Observatory of Japan. All
image reduction is performed with IRAF\footnote{IRAF is the Image
Analysis and Reduction Facility made available to the astronomical
community by the National Optical Astronomy Observatories, which are
operated by AURA, Inc., under contract with the U.S. National Science
Foundation.}. Bias and dark counts are subtracted from all images. For
the $V$ and $I$ band images, the flat fields are obtained from the
median of dome-flat frames. For the $J$ band images, the flat fields
are obtained from the median of sky frames acquired before and after
the object frames. The object frames are flat-fielded and then
sky-subtracted. The field of view of the cameras is large enough to
determine the sky background around the galaxies in the image. The sky
background counts are estimated on the object frames with SPIRAL
package (\cite{ichikawa1987}) installed in IRAF. Finally, the images
are rotated to let the major axis of the galaxy horizontal to the
bottom.  Position angles (PA) taken from RC3 are used to rotate the
image for many galaxies. However, we find that the PAs of RC3 are
sometimes not suitable for some galaxies from an inspection of model
fitting as mentioned in the next section. Therefore we try the
following PAs for the galaxy to fit better with the surface brightness
model: 0, 30, 60, 90, 120 and 150 degree plus PA of RC3. The best PAs
are ascribed in section 4.3.

The images are photometrically calibrated with the aperture magnitudes
taken from \citet{devaucouleurs1988}. Typical photometric errors of
the calibration are 0.05 mag for the $V$ and $I$ bands, and 0.08 mag
for the $J$ band.

\section{Analyses for The Galaxy Images}

\subsection{Method of Surface Brightness Fitting}

First, the galaxy images are decomposed into bulge and disk
components. Our decomposition method follows \citet{byun1995}. Here
the contour of the surface brightness is assumed to be axisymmetric
elliptical for both bulge and disk. The centers and the position
angles of the bulge and disk are assumed to be common. The
2-dimensional version of generalized model of the surface brightness
(\cite{sersic1968}) is
\begin{eqnarray}
I(r)=I_e \exp[ -b_n ((r/r_e)^{\beta}-1) ],\nonumber\\
r=\sqrt{x^2+(y/(b/a))^2},
\end{eqnarray}
where $\beta$ is the slope of exponent (according to
\cite{mollenhoff2001}, we use $\beta$ instead of traditional $1/n$),
$I_e$ the effective luminosity in intensity unit, $r$ the radius from
the galactic center, $r_e$ the effective radius, $b_n$ a constant
relating $I_e$ and $r_e$ (see \cite{moriondo1998a}), $x$ and $y$
distances from the center along major and minor axes, respectively,
and $b/a$ the axial ratio. The $r$ of bulge and disk are hereafter
denoted as $r_b$ and $r_d$, respectively.  The $b/a$ of bulge and disk
are hereafter denoted as $(b/a)_b$ and $(b/a)_d$,
respectively. $\mu_e$ in mag arcsec$^{-2}$ is also used instead of the
$I_e$.  For disk, we assume $\beta$=1 (exponential law)
(\cite{freeman1970}). Thus the surface brightness of disk is
\begin{eqnarray}
I_d(r)=I_0 \exp[ -(r_d/r_h) ],\nonumber\\
r_d=\sqrt{x^2+(y/(b/a)_d)^2},
\end{eqnarray}
where $I_0$ and $r_h$ are the central luminosity and the scale length,
respectively. $\mu_0$ in mag arcsec$^{-2}$ is also used instead of the
$I_0$.

For bulge, various values of the slope $\beta=1/n$ have been reported
(e.g., \cite{caon1993}; \cite{andredakis1994}; \cite{andredakis1995};
\cite{dejong1996a}; \cite{seigar1998}; \cite{moriondo1998a};
\cite{mollenhoff2001}). In general, $\beta$ has thought to be within
the range of 1 (exponential law) to 1/4 (de Vaucouleurs' law)
(\cite{devaucouleurs1953}). For $\beta=1$, 

\begin{eqnarray}
I(r)=I_e \exp[ -1.68 ((r_b/r_e)^{\beta}-1) ],\nonumber\\
r_b=\sqrt{x^2+(y/(b/a)_b)^2},
\end{eqnarray}
For $\beta=1/4$, it is 
\begin{equation}
I(r)=I_e \exp[ -7.67 ((r_b/r_e)^{\beta}-1) ].
\end{equation}
If the $\beta$ is the free parameter in the decomposition, it is not
generally well determined, because the point spread function (PSF) and
the disk parameters may affect the estimating $\beta$ (e.g.,
\cite{dejong1996a}; \cite{moriondo1998a}; \cite{balcells2003}).
\citet{dejong1996a} and other recent studies claimed that about more
than fifty percent bulges in nearby galaxies are fitted with
exponential law rather than de Vaucouleurs' law. We therefore fixed the
value of $\beta$. At first we try $\beta=1/4$, 1/3, 1/2, and 1.  We
find that the fitted value of $\beta$ is 1 or 1/2 for most of our
sample galaxies. Then we add $\beta=2/3=0.67$ and $\beta=3/4=0.75$
models to examine carefully the bulge shape between $\beta=1/2$ and
$\beta=1$. As the results, some galaxies (NGC 3198, 3521, 4258, 4548,
4736, 5907 and 6946) are well fitted using the $\beta=2/3$ or
$\beta=3/4$ bulge models rather than the $\beta=1/2$ or $\beta=1$.

NGC 1068, 1808, 2841, 3031, 3079, 4192, 4258, 4303, 4548, 4565, 4569
and 5194 are classified as Seyfert galaxies. Some other galaxies also
have a bright central point source, however they are not classified as
Active Galactic Nuclei (AGN). Hence the nuclear parameters (central
luminosity of the point source $I_{nuc}$, or $M_{nuc}$ in magnitude
unit) are added into the bulge/disk decomposition program. The total
surface brightness of a galaxy is sum of $I_b$, $I_d$ and $I_{nuc}$.
In practice, the extra luminosity $I_{nuc}$ turns out to be negative
value for some galaxies. When the nucleus luminosity is negative, the
galaxy is fitted with only bulge and disk parameters.  To simplify,
spiral arms, bar and ring structures are not included in our model.

The observed images are convolved with seeing and instrument point
spread function (PSF). Hence it is required that the observed images
are deconvolved with the PSF or the model images are
convolved. However, the deconvolution may add the artificial noise and
may change the profile of the galaxy (especially for
bulge). Therefore, instead of the deconvolution for the observed
image, we convolved the model surface brightness.

In general, the higher signal-to-noise ratio ($S/N$) and the larger
number of pixels give the better analysis. The distances of $x$ and
$y$ from the galactic center having $S/N>10$ are measured and the
pixels within the rectangle are used in the analysis. We found
empirically that this method gives better fitting.

To obtain the parameters, chi-square ($\chi^2$) is defined as

\begin{equation}
\chi^2=\sum_{i}\frac{(I_{obs}(x,y)-I_m(x,y))^2}{\sigma^2_i},
\end{equation}
where $I_{obs}(x,y)$ is the observed value of the galaxy surface
brightness, and $I_m(x,y)=I_b(x,y)+I_d(x,y)(+I_{nuc}(x,y))$ is
convolved with Moffat function (\cite{moffat1969}),

\begin{equation}
{\rm PSF}(r)=(\beta-1)[1+(r/\alpha)^2]^{-\beta}/\pi\alpha^2,
\end{equation}
\begin{equation} 
{\rm FWHM}=2\alpha\sqrt{2^{1/\beta}-1}, 
\end{equation}

where FWHM and $\beta$ (this is not the bulge slope) are taken from
the average values for those of a few stars in the frame calculated
from the ``imexam'' task of IRAF. Being assumed Poisson distribution,
the uncertainty ($\sigma_i$) is set to the square root of
$(I_{obs}(x,y)+I_{sky}(x,y)+(I_{read}/\alpha)^2)/\alpha$, where
$I_{sky}$ is the sky level, $I_{read}$ the read noise, and $\alpha$
the conversion factor. We confirm that this $\sigma_i$ gives the
smallest scattering in the comparison of input-output parameters for
our simulated images as discussed in section 3.2.  The parameters in
the model are obtained by a nonlinear least square fitting, which
minimizes the $\chi^2$. Our calculation is based on Marquardt method
(\cite{press1988}).  Various models including $\beta=1/n$ and PA are
tried for each galaxy, and the best model that $\chi^2$ is the smaller
than other models is found.

Although the $\chi^2$ is useful to find the better model for one
galaxy, they are not useful to compare the error of fitting between
sample galaxies, because the $\chi^2$ may become large or small due to
the adopted weighting factor $1/\sigma^2_i$, as \citet{baggett1998}
have pointed out. To compare the error of fitting between galaxies, we
calculate the total residual
($\sqrt{\sum_{i}(I_{obs}(x,y)-I_m(x,y))^2}$) in magnitude unit. They
are tabulated in table 3-5 and are about 0.1-0.3 mag.

We next correct the obtained bulge $\mu_e$ and disk $\mu_0$ with the
Galactic extinction. The extinction is taken from the table of NED
(NASA/IPAC Extragalactic Database)\footnote{The NASA/IPAC
Extragalactic Database (NED) is operated by Jet Propulsion Laboratory,
California Institute of Technology, under contract with the
U.S. National Aeronautics and Space Administration.}, which is
calculated from the data of \citet{schlegel1998}.  The total model
magnitudes of bulge, disk and the total galaxy (bulge + disk +
nucleus) are calculated by integrating the model surface brightness of
each component from center to infinity for each band.  The colors of
bulge and disk are calculated from the model magnitudes of $V$, $I$
and $J$ bands.

\subsection{Accuracy of The Surface Brightness Fitting}

We construct model images and then decompose them to examine how well
the parameters are reproduced by our analysis. We make 300 models for
each $\beta=1/4$ (de Vaucouleurs), $1/3$, $1/2$, $2/3$, $3/4$, and $1$
(exponential) bulge $+$ exponential disk $+$ nucleus. The values of
the six parameters are selected at random. Three parameters are set to
the following range: $0.3<(b/a)_b<1.0$, $0.1<(b/a)_d<0.8$ and
$30<r_h<150~{\rm arcsec}$. Also the disk $\mu_0$ are set to
$18<\mu_{0,V}<22$, $17<\mu_{0,I}<21$, and $16<\mu_{0,J}<20~{\rm
mag~arcsec^{-2}}$ for $V$, $I$, $J$ band, respectively. Also the
nuclear magnitude $M_{nuc}$ are set to $12<M_{nuc,V}<16$,
$11<M_{nuc,I}<15$, and $10<M_{nuc,J}<14~{\rm mag~arcsec^{-2}}$ for
$V$, $I$, $J$ band, respectively.

Since $\mu_{e}$ and $r_{e}$ are inversely proportional to the slope
$\beta$ (cf., \cite{dejong1996a}; \cite{moriondo1998a}), the model
$\mu_{e}$ and $r_{e}$ should be set to suitable ranges for the assumed
$\beta$. To simplify, $\mu_e$ and $r_e$ for $\beta=1/4$ and
$\beta=1/3$ are set to $19<\mu_{e,V}<23$, $18<\mu_{e,I}<22~$,
$17<\mu_{e,J}<21~{\rm mag~arcsec^{-2}}$ and $10<r_e<100~{\rm arcsec}$,
while those for $\beta=1/2$, $2/3$, $3/4$, and 1 are set to
$18<\mu_{e,V}<22$, $17<\mu_{e,I}<21~$, $16<\mu_{e,J}<20~{\rm
mag~arcsec^{-2}}$ and $5<r_e<50~{\rm arcsec}$. Those parameter ranges
are typical in our sample galaxies.

Here we show the results using the instrument features of 1kCCD and
KONIC. The results for 2kCCD camera are almost the same as for
1kCCD. To simulate the images observed with 1kCCD camera or KONIC, the
exposure times are set to 300s for $V$ and $I$ band and 1800s for $J$
band. The specifications of 1kCCD camera for $V$ and $I$ bands and
those of KONIC for $J$ band are input to the model images. The {\it V}
and $I$ band images are convolved with Moffat function. The FWHM is
assumed to be 3.8 arcsec, and the parameter $\beta$ is set to 4.0,
which are average values in our galaxies. The $J$ band images are
convolved with a special PSF for KONIC\footnote{The data is available
at
http://www.ioa.s.u-tokyo.ac.jp/kisohp/INSTRUMENTS/konic/psf2.gif.}. Then,
the background noise according to the exposure time and the
performance of the detectors are added to the convolved
images. Finally those simulated image is decomposed into bulge, disk
and nucleus to obtain the seven parameters. Then the input parameters
are compared with the outputs.

\begin{figure*}
  \begin{center} \FigureFile(170mm,170mm){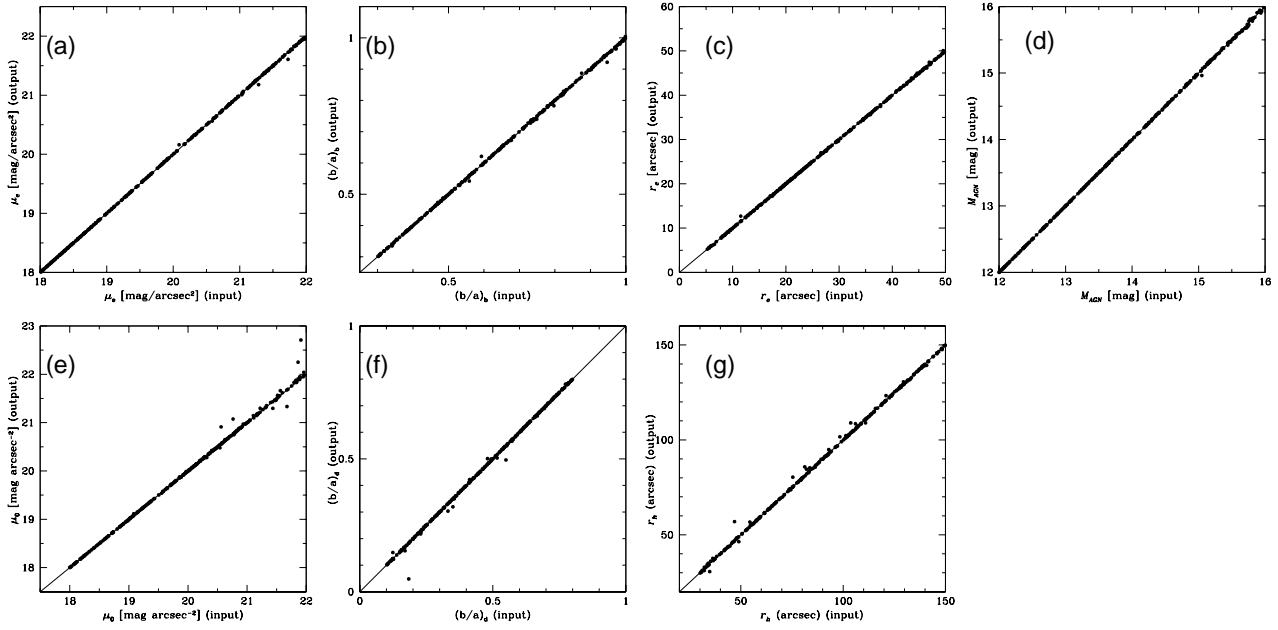} \end{center}
   \caption{The result of the two-dimensional decomposition with
   artificial galaxies for $\beta=1$ bulge model and $V$ band. The
   solid lines in each panel show the equality of input and output
   parameters. (a) bulge effective luminosity $\mu_e$. (b) bulge axial
   ratio $(b/a)_b$. (c) bulge effective radius $r_e$. (d) Nucleus
   magnitude $M_{nuc}$. (e) disk central luminosity $\mu_0$. (f) disk
   axial ratio $(b/a)_d$. (g) disk scale length $r_h$.}  \label{fig.1}
\end{figure*}

\begin{figure*}
  \begin{center} \FigureFile(170mm,170mm){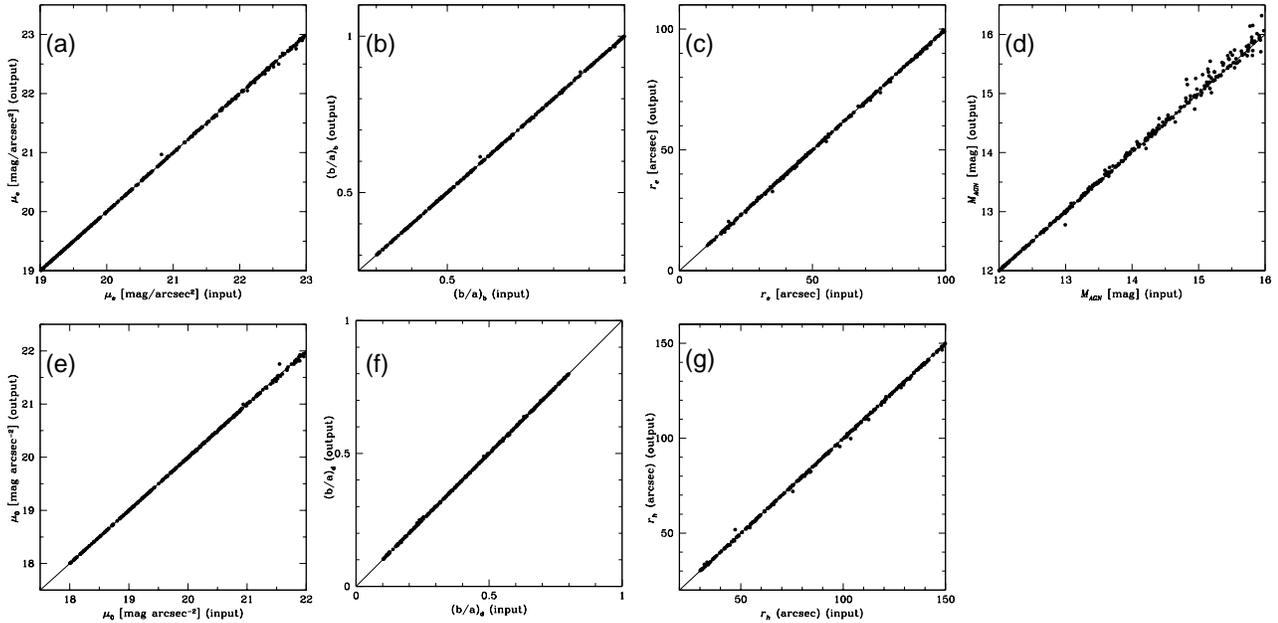} \end{center}
   \caption{The same as for figure \ref{fig.1}, but for $\beta=1/4$
   bulge model and $V$ band. }  \label{fig.2} \end{figure*}

\begin{figure*}
  \begin{center} \FigureFile(170mm,170mm){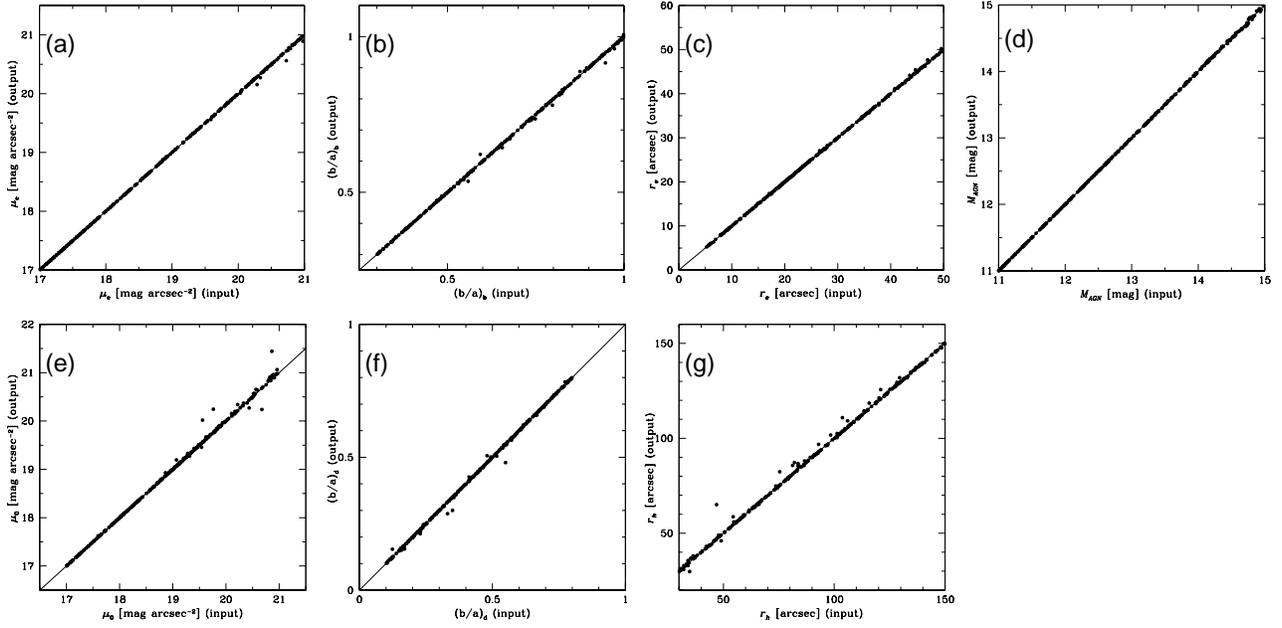} \end{center}
    \caption{The same as for figure \ref{fig.1}, but for $I$ band. }
    \label{fig.3} \end{figure*}

\begin{figure*}
  \begin{center}
    \FigureFile(170mm,170mm){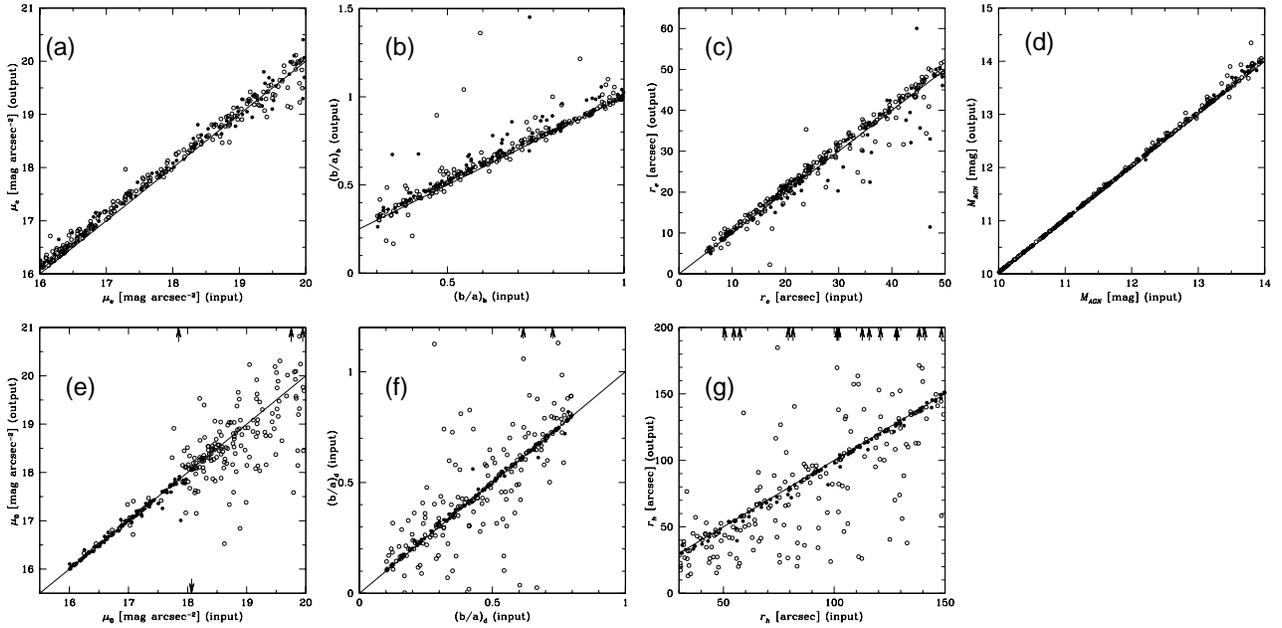}
  \end{center}
  \caption{The same as for figure 1, but for $J$ band. The filled and
 open circles denote the models within the range of $0.01<B/T<0.4$ and
 $\mu_{0,J}<18.0$ and the models outside the range,
 respectively. The arrow shows the data located outside the figure.}
 \label{fig.4} 
\end{figure*}

{}From figures \ref{fig.1}-\ref{fig.4}, we can look how well the
parameters are reproduced. Here we show the results of $\beta=1$ and
$\beta=1/4$ for $V$ band (figure \ref{fig.1}-\ref{fig.2}), $\beta=1$
for $I$ band (figure \ref{fig.3}), and $\beta=1$ for $J$ band (figure
\ref{fig.4}). Figure \ref{fig.1}-\ref{fig.3} indicate that the
simulated images are fitted with the models in high accuracy. Other
models ($\beta=1/3,1/2,2/3,3/4$) are also well fitted, however they
are not illustrated. We find that the $1\sigma$ errors of all
parameters for the $V$ and $I$ band images are less than 1\% of the
original values expect for the $M_{nuc}$ at the $\beta=1/4$ models. On
the other hand, the errors are relatively large in the $J$ band
(figure \ref{fig.4}). The large amount of background noise should
cause the fitting error for $J$ band. However the scattering is
relatively small within a limited range of parameter values;
$0.01<B/T<0.4$ and $\mu_{0,J}<18$, which is denoted as black circles
in figure \ref{fig.4}. Therefore we only use the fitting results
within the permitted range. The 1$\sigma$ errors in the range for the
$J$ band are 0.17 mag, 0.06, 5.7 arcsec, 0.16 mag, 0.02, 5.0 arcsec,
0.01 mag for bulge $\mu_e$, $(b/a)_b$, $r_e$, disk
$\mu_0$, $(b/a)_d$, $r_h$, and $M_{nuc}$, respectively.

\subsection{Model of Rotation Curve}

After the surface brightness decomposition of the galaxies images, the
rotation curves of \citet{sofue1997} and \citet{sofue2003} are
decomposed into the contribution from the bulge, disk and halo
component to obtain those $M/L$s.  On the assumption of the constant
$M/L$, the mass surface density distribution $M(r)$ for bulge or disk
component is expressed as
\begin{equation}
M(r)=(M/L) \times I(r),
\end{equation}
where $I(r)$ is the luminosity density for bulge or disk in solar
luminosity unit. The unit of $M/L$ is $(M/L)_\odot$ (ratio of solar mass
to solar luminosity).

The circular velocity for ellipsoidal bulge is given by
(\cite{binney1987}):
\begin{equation}
v_b^2(r)=4{\pi}G(M/L)_{b}\sqrt{1-\epsilon^2}\int_{0}^{r}\frac{j_{b}(a)a^{2}da}{\sqrt{r^{2}-a^{2}\epsilon^{2}}},
\end{equation}
where $(M/L)_b$ the bulge $M/L$, $j_b(a)$ the luminosity density at
distance $a$ from the center in the equatorial plane, and
$\epsilon=\sqrt{1-b^2/a^2}$ the constant intrinsic eccentricity.  To
obtain the $j_b(a)$, we consider the ``strip brightness'' $S(a)$,
defined as the integral of the bulge brightness distribution along a
path orthogonal to the line of nodes, at distance $a$ from the center
(Moriondo et al. 1998b). Then
\begin{equation}
\frac{dS}{da}=-2{\pi}\sqrt{1-\epsilon^2}aj_b(a).
\end{equation}

On the other hand, the circular velocity for exponential disk is given
by (\cite{freeman1970}):
\begin{equation}
v^2_d(r)=4{\pi}G{\Sigma}(0)r_{h}y^{2}[I_0(y)K_0(y)-I_1(y)K_1(y)],
\end{equation}
where $y=r/2r_h$, $I_n(y)$ and $K_n(y)$ are the modified Bessel
functions of first and second kind, respectively,
$\Sigma(0)=(M/L)_dI_0$ the central disk surface density, and $(M/L)_d$
the disk $M/L$.

The mass of the nucleus is about $10^7M_{\odot}$ for NGC 1068
(\cite{greenhill1996}) and generally smaller than $10^6-10^7M_{\odot}$
for other spirals (\cite{salucci2000}), which influence hardly the
rotation curves at least the 15 or 2-4 arcsec resolution of
\citet{sofue1997} or \citet{sofue2003} rotation curves. Therefore the
contribution of nucleus on rotation curve is ignored in this paper.

The observed rotation curves may be superposed by non-circular motions
induced by bar. The bar-influenced gas flow at a side-on view gives
lower rotation velocity, whereas that at an end-on view give higher
rotation velocity (\cite{sofue1997}). If the effect of bar-influenced
gas flow is large, most of rotation curves will indicate lower and
more gradually rising features, because the probability of looking bar
at an end-on view is lower than that at an side-on view
(\cite{sofue1997}).  However, \citet{sofue1997} reported that most of
rotation curves have the steep rise features and there are no
significant difference in the rotation curves between the barred
galaxies and the normal spiral galaxies. Hence the
steep rise feature should be due to mass concentration within the
bulge at zero-th order approximation (\cite{sofue1997}). Therefore the
effects of bar is ignored.

There is a massive dark matter (halo) in galaxies. The halo produces
flat or nearly flat rotation curves in most galaxies (e.g.,
\cite{barnaby1994}). The pseudo-isothermal sphere model possesses this
property, and its density distribution is given by (\cite{kent1986}):
\begin{equation}
\rho(r)=\rho_{0,h}/[1+(r/R_h)^2] = \sigma_h/[2\pi G(r^2+R_h^2)],
\end{equation}
where $\rho_{0,h}$ the halo's central mass density, $R_h$ the halo core
radius, and $\sigma_h$ the velocity dispersion. This
distribution gives the rotation curve of

\begin{equation}
v_h(r)^2=2\sigma_h^2[1-(R_h/r) \tan^{-1}(r/R_h)].
\end{equation}

The dark halo parameters are strongly affected by the observed maximum
radius of rotation curve. In this paper we consider mainly $M/L$ of
luminous matter, thus the discussion in the reliability of halo
parameters is beyond our scope. The influence of dark halo blending on
the maximum bulge plus disk solutions is discussed in the next
section.

The gravitational potential, $\phi$, of a spiral galaxy is a
superposition of the potentials generated by its individual mass
components (\cite{kent1986}; \cite{barnaby1994}),
\begin{equation}
\phi = \phi_b + \phi_d + \phi_h.
\end{equation}

Thus we get an expression for the model rotation velocity
\begin{equation}
v(r)^2=v_b(r)^2 + v_d(r)^2 + v_h(r)^2.
\end{equation}

We use again Marquardt method to fit rotation curves. Here the
uncertainty of observed rotation curve is set to constant over the
full range, because the error size of the observed rotation curve is
nearly constant from the center to the end. We input the uncertainty
of 15 [km s$^{-1}$]/$\sin i$ (\cite{sofue1997}; \cite{sofue2003}) to
rotation curve fitting, where $i$ is the inclination of the
galaxy. The $i$ are calculated with $i=\cos^{-1}(1/{\rm R_{25}})$,
where ${\rm R_{25}}$ are axial ratios taken from RC3.

The parameters used in the rotation curve fitting are $(M/L)_b$,
$(M/L)_d$, $\sigma_h$, and $R_h$. The solution of these four
parameters obtained simultaneously is called "full solution". However,
in practice, the full solution is often singular because the observed
rotation curves do not extend far enough to decouple the
interdependence of the disk $M/L$ ratio and halo parameters
(\cite{kent1986}). Hence \citet{albada1985} and \citet{kent1986}
devised the method restricting parameters and obtaining the solutions
step by step. The solutions obtained with the method is called
"maximum bulge plus disk solution" or "maximum disk solution". We also
adopt the method in our analysis. In this method we assume that
rotation curve can be fitted by only visible matter within the radius
of two times of disk scale length ($r_h$), which gives maximum $v_d$
and thus maximum $(M/L)_d$. Thus two parameters ($(M/L)_b$ and
$(M/L)_d$) are determined within $2r_h$. Next, holding the resulting
values, we obtain the remaining parameters ($\sigma_h$ and $R_h$) by
fitting the full range of rotation curve.

If the observed rotation curve has the slowly rising feature compared
to the luminosity concentration of bulge and is well fitted by only
disk contribution, the $(M/L)_b$ tends to be nearly zero or
negative. In fact, we obtain negative $(M/L)_b$ for NGC 3198 in the
$I$ band. The effect of beam smearing of rotation curve would cause
the slowly rising feature. Since the negative values of $M/L$s are
nonsense, we restrict the parameters to positive in the fitting.

\subsection{Accuracy of The Rotation Curve Fitting}

In order to examine an accuracy of the rotation curve fitting, we
construct 300 model rotation curves and then apply our method to the
models.  The distances are set to 10 Mpc. The radii of the rotation
curves are fixed to 500 arcsec, which corresponds to 24 kpc at the
distance of 10 Mpc. These are the typical observations of
\citet{sofue1997}. We select the parameters at random within the
following range: $0.5<(M/L)_b<3.0$, $0.5<(M/L)_d<3.0$,
$100<\sigma_h<150 ~{\rm km~s^{-1}}$, $100<R_h<500 ~{\rm arcsec}$. The
ranges of luminosity parameters ($\mu_e$, $\mu_0$ and so on for each
$\beta$, $V$, $I$ and $J$) input in the simulation are the same as
above section. The ranges of $(M/L)_b$, $(M/L)_d$, $\sigma_h$ and
$R_h$ are taken from the typical values in \citet{kent1986}. Here we
show the result with the $\mu_e$ and $\mu_0$ for $\beta=1$ and $I$
band. Other cases of $\beta$ or bands are also tried, and the similar
results are obtained. Using these parameters, the dark-to-luminous
(bulge $+$ disk) mass ratio of model galaxy is within the range of
0.07 (luminous matter dominating) to 32.2 (dark matter dominating)
(typically about 3), and the total $M/L_I$ ratio of the galaxy is
within the range of 1.0 to 69.1 (typically about 6). These values are
generally consistent with those of previous authors (e.g.,
\cite{kent1986}, \yearcite{kent1987}).

The rotation curves in \citet{sofue1997} and \citet{sofue2003} were
observed with the angular resolution of 15 arcsec and 2-4 arcsec,
respectively. Hence the model rotation curves are convolved with
Gaussian with FWHM of 15 arcsec or 3 arcsec. Here we show the result
of the case of 15 arcsec resolution. Finally the simulated rotation
curves are fitted to the velocity model with the maximum bulge and
disk method.

The radius from the galactic center used in the maximum disk method
affects the accuracy of fitting. In general, a large radius would
bring a contamination of dark halo, while a small radius would bring a
confusion of bulge and disk. Thus we examine radii of 1, 2 and 3 times
of disk $r_h$. We confirm that 2 $r_h$, which was used also in
\citet{moriondo1998b}, gives the best fitting. Therefore we adopt 2
$r_h$ in the following discussion.

\begin{figure*}
  \begin{center}
   \FigureFile(180mm,180mm){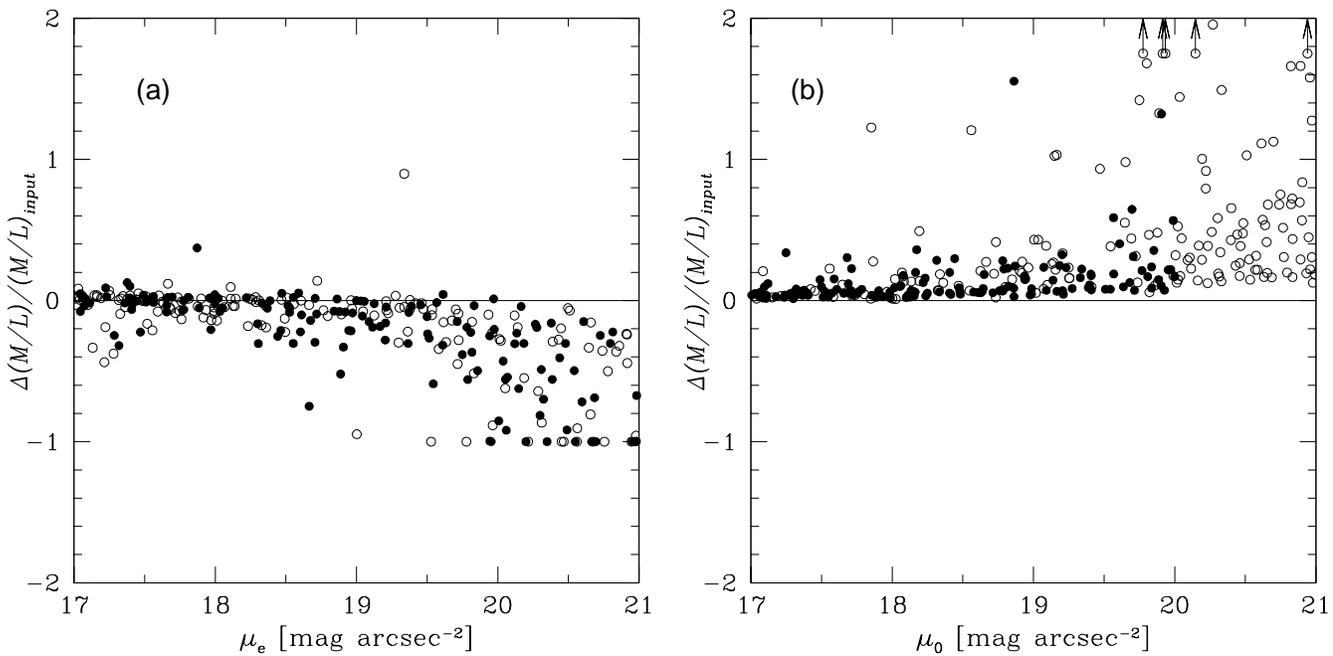} \end{center}
    \caption{The result of the rotation curve decomposition with
    artificial galaxies for $I$ band. (a) : The comparison of the
    relative residuals ((output-input)/input) of bulge $M/L$ versus
    $\mu_e$. The filled and open circles denote the models within the
    range of $0.01<B/T<0.5$ and $\mu_{0,I}<20$ and the models outside
    the range, respectively. (b) : The same as (a), but for disk $M/L$
    versus $\mu_0$.}  \label{fig.5}
\end{figure*}

{}Figure \ref{fig.5} shows the comparisons of the relative residuals
((output$-$input)/input) with $\mu_e$ or $\mu_0$. The systematic error
for bulge $M/L$ against $\mu_e$ and disk $M/L$ against $\mu_0$ are
clearly seen in the figure. The bulge error is mainly caused by the
disk contamination, while the disk error is mainly caused by the dark
halo contamination. We find empirically that the errors are relatively
small for the range of $\mu_{e,I}<19.5$, $\mu_{0,I}<19.5$
($\mu_{e,V}<20.5$, $\mu_{0,V}<20.5$ for $V$ band, $\mu_{e,J}<18.5$,
$\mu_{0,J}<18.5$ for $J$ band), and $0.01<B/T<0.5$. We therefore
reject the results outside the above ranges. For about 6 \% of the
models, the output values of $M/L$s are smaller than 0.1, which are
clearly out of the range predicted from galaxy formation
models. Therefore we adopt only the models of output $M/L>0.1$ for
estimating the error of $M/L$. Using the result of the examination, we
correct the fitting results of bulge and disk $M/L$s for sample
galaxies according to $\mu_e$ and $\mu_0$, respectively. The
correction of systematic error of $M/L$s for bulge and disk are not
considerably large within the reliable parameter ranges: about factor
of 1.0-1.3 and 0.9-1.0 of originally obtained values,
respectively. The scatterings of bulge and disk $M/L$s within the
ranges in our simulation are about 10-15\% and 6-8\%, respectively. In
practice, since the errors of surface brightness parameters and the
residual $\chi^2$ are added, the errors of $M/L$ for observed galaxies
are somewhat enlarged; about 15-30\% for both of bulge and disk $M/L$.

\section{Results}

\subsection{Fitting Result of Surface Brightness and Rotation Curves}

\begin{figure*}
  \begin{center} \FigureFile(180mm,180mm){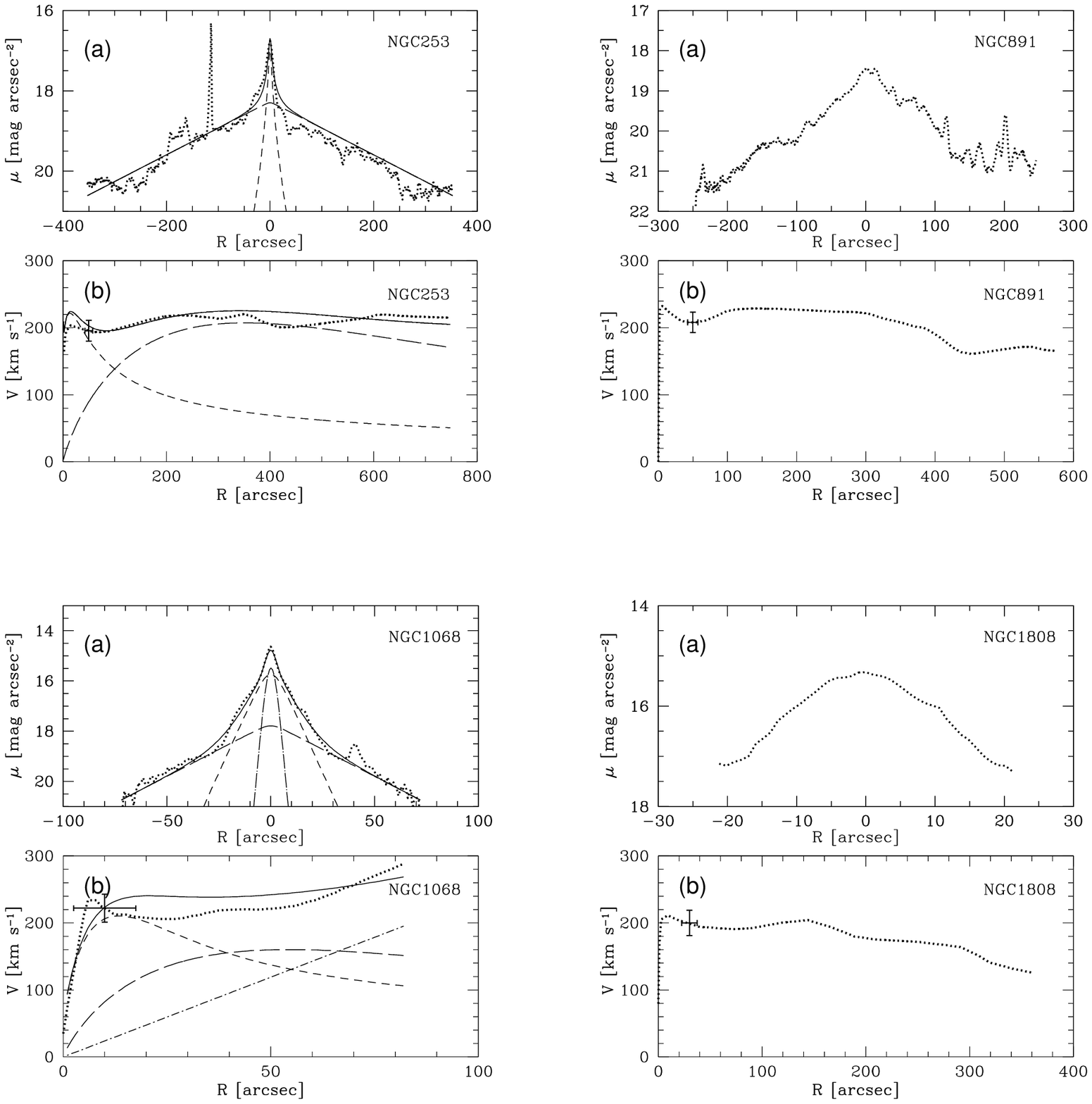} \end{center}
    \caption{(a) observed and model surface brightness profiles along
    major axis. The dotted lines, solid lines, short-dashed lines,
    long-dashed lines and dot-long dashed lines represent the observed
    profile, the model profile, bulge, disk and nuclear contributions,
    respectively. The profiles of NGC 1808, NGC 3628, NGC 4303 and NGC
    4631 are those of $J$ band images, and the remainder are those of
    $I$ band images. (b) observed and model rotation curves. The
    symbols are the same as for (a). The contribution from dark halo
    is shown as the dot-dashed lines. The error bars for the rotation
    curves represent the angular resolution of 15 arcsec
    (\cite{sofue1997}) or 3 arcsec (\cite{sofue2003}), and the
    velocity resolution of 15 ${\rm km s^{-1}}/\sin(i)$ for both of
    \citet{sofue1997} and \citet{sofue2003}, where $i$ is the
    inclination of galaxy. } \label{fig.6.1}
\end{figure*}

Figure \ref{fig.6.1} shows the luminosity profiles and rotation curves
of sample galaxies. The results of fittings for surface brightness and
rotation curve are listed in tables 3-5 and 7-9, respectively.

The surface brightness parameters for NGC 1808 ($J$) and 4536 ($J$)
are not obtained because of poor weather condition. The edge-on galaxy
NGC 891 is also not well fitted to any model in all bands because of
the dust lane. The parameters for NGC 4321 and 4535 in the $J$ band
are unreliable, because they are dimmer than the reliable range of
disk central luminosity $\mu_{0,J}$. The parameters for NGC 2903,
3198, 4258 and 4736 in the $J$ band may be also unreliable, because
the image quality ($S/N$) are relatively low and the values of $r_e$
or $r_h$ in the $J$ band for these galaxies are considerably different
from those of $V$ and $I$ band. We exclude these data in the
discussion. The image of nucleus region for NGC 4736 ($I$) is
saturated, hence the parameters may be somewhat inaccurate,
however the residual of fitting is small. The results of surface
brightness fitting for 22 ($V$), 23 ($I$) and 10 ($J$) galaxies are
reliable and are used in the discussion of the structural parameter,
color and magnitude.

The $\chi^2$ values in the surface brightness fitting turn out to be
almost the same among various $\beta=1/n$ models of bulge. Thus the
$\beta$ is not well determined only by the surface
brightness. However, the $\chi^2$ value in the rotation curve fitting
clearly depends on $\beta$, because the central rising and peak of
rotation curve is sensitive to the bulge $\beta$. Consequently we
obtain the $\beta$ by the result of rotation curve fitting. In fact,
the central rising part of rotation curve observed in
\citet{sofue1997} and \citet{sofue2003} are well reproduced by the
exponential bulge in many galaxies.  The fact that the $\beta\ge1/2$
model is more suitable than $\beta<1/2$ model has been reported by
other authors (e.g., \cite{dejong1996b}, \cite{mollenhoff2001},
\cite{graham2001}, \cite{macarthur2003}, \cite{hunt2004}). Our results
agree with these recent studies.

When bulge $\mu_e$ and/or disk $\mu_0$ are larger (fainter) than the
reliable ranges, the $M/L$s of bulge or disk are also
unreliable. Furthermore, the bulge $M/L$ (and color) in the galaxy of
inclination $i>75$ degree would be influenced by dust extinction and
thus unreliable, as mentioned in section 4.4. These data are denoted
as colon in the tables 7-9. The bulge $M/L$ for 13 ($V$), 14 ($I$) and
5 ($J$) galaxies and the disk $M/L$ for 17 ($V$), 21 ($I$) and 11
($J$) galaxies are reliable and are used in the discussion.

Indeed, the rotation curves are not perfectly fitted with the model,
especially for the inner region of NGC 1068, NGC 2841, NGC 3031, NGC
4548 and NGC 4569. There would be some reasons for the
discrepancy. One reason may be the gravitation of massive components
(e.g., inner bar and ring) other than bulge and disk around the
galactic center. It would be appropriate for NGC 1068 and NGC 4548;
the former has a ring and another complex structure in the inner
region (RSA(rs)b), and the latter is classified as the barred galaxy
(SB(rs)b). Another reason would be the low resolution of 15 arcsec for
CO rotation curves. It may be valid partly for NGC 1068, NGC 2841 and
NGC 3031. The reason of discrepancy for NGC 4569 rotation curve around
the galactic center, of which resolution is 4 arcsec
(\cite{sofue2003}), remains unknown in this paper. The more careful and
complex method of model fitting for rotation curve should be studied
in the future.

The halo parameters are significantly affected by the observed radius
of rotation curve from the galactic center (cf. \cite{kent1986};
\cite{moriondo1998b}). If the radius does not extend far enough to
decompose into luminous matter and dark matter, the rotation curve can
be fitted with only bulge and disk, and the halo parameters are not
well determined. In fact, the halo parameters for some galaxies are
not determined or implausible. Since we focus on only $M/L$s for
bulges and disks in this paper, the halo parameters are not used in
the following discussion.

\subsection{The Trend of Bulge and Disk Scale Parameters}

We compare the scale length and bulge effective radius in three bands
to check the validity of the fitting.  Figure \ref{fig.7} and
\ref{fig.8} show the trend of $r_e$ and $r_h$ on the bands,
respectively. The values are normalized for those of $V$ band. Only 13
galaxies (NGC 253, 2841, 2903, 3031, 3079, 3198, 3521, 4258, 4321,
4535, 5907, 6946 and 7331) are decomposed for all ($V$, $I$, $J$)
bands and plotted in the figures.  Those of $J$ band for NGC 2903,
3198, 4258, 4321 and 4535 are unreliable as described in the previous
section and are connected by dashed lines in the figures. There is no
significant trend between the bulge effective radius $r_e$ and the
band in figure \ref{fig.7}, while \citet{macarthur2003} reported that
$r_e$ increases with longer wavelength band, however the scattering is
large. On the other hand, the disk scale length $r_h$ generally
decreases from $V$ to $J$ in figure \ref{fig.8}. The trend has been
reported in \citet{mollenhoff2004}.  The decreasing $r_h$ toward the
redder bands would be caused by the higher concentration of old stars in
the central region and/or an increasing of star formation in the outer
disk (e.g., \cite{macarthur2003}; \cite{mollenhoff2004}). More sample
is necessary to discuss the correlation.

\subsection{The Position Angles and Axial Ratios}

To simplify, the PA of galaxy is not the fitting parameter but is
fixed to 0, 30, 60, 90, 120 and 150 degree plus the RC3 data in this
paper. Each galaxy image rotated with these PAs is fitted with the
surface brightness model to decide the best PA from the smallest
$\chi^2$. As the result, the PAs for many galaxies are 0 plus RC3's
PA. That is to say, the RC3 data are suitable. The PAs for some nearly
face-on galaxies do not agree with RC3 data, however. We adopt the
following PAs for some galaxies: RC3's PA plus 30 degrees for NGC
1068, plus 60 degree for NGC 4303, plus 90 degrees for NGC 4321 and
NGC 4548, 120 degrees for NGC 5194 and 60 degrees for NGC 6946.  The
discrepancy of obtained PA between our analysis and RC3 for some
face-on galaxies would be due to the following reason: The parameters
are derived from the luminosity-weighted fitting in inner region of
$S/N>10$ (corresponding to about 23 $V$-magnitude) in this paper,
while the ratio in RC3 is measured at the contour of 25
$B$-magnitude. That is, our results are obtained from inner disk, in
which the bar or spiral arms sometimes dominate, while the PA in RC3
is measured at outer disk. These structure are apparent in face-on
galaxies and often cause the discrepancy of the PAs.

We compare the model disk axial ratio $(b/a)_d$ obtained in surface
brightness fitting with the catalogue axial ratios $R_{25}$ taken from
RC3. The comparison is shown in figure \ref{fig.9}.  The model axial
ratio is generally consistent with $R_{25}$ in RC3.

The bulge axial ratio $(b/a)_b$ for some galaxies are larger than 1.0
in table 3-5. That is, the major- and minor-axis for these bulge is
reverse against the disk. In other words, the $(b/a)_b>1$ indicates
that the PA of bulge is nearly vertical to that of disk. The
$(b/a)_b>1$ would be due to a distorted structure of bulge. If we
construct the model including parameters of individual PAs for both
bulge and disk, the $(b/a)_b>1$ would not happen. To simplify, we do
not use the parameters of individual PAs for bulge and disk in model,
however.  When $(b/a)_b$ is larger than 1, the intrinsic eccentricity
$\epsilon=\sqrt{1-(b/a)_b^2}$ cannot be calculated and therefore the
$(M/L)_b$ also cannot be obtained. We apply $\epsilon=0.0$ (assuming
spherical bulge) to calculate the $M/L_b$ if the $\epsilon$ cannot be
calculated. Fortunately, the dependence of $(M/L)_b$ on $\epsilon$ is
relatively small; smaller than $10\%$ within
$0\le\epsilon<1$. Therefore the error is not so large even if we
assume the spherical bulge, and therefore the uncertainty of bulge PA
does not considerably affect the resulting $M/L$.

\begin{table*}
\begin{center}
 \caption{\bf Results of $V$ band surface brightness fitting.}
\begin{tabular}{c ccc c ccc ccc c c c}    
\hline\hline
 Name &  $\mu_e$ & $(b/a)_b$ & $r_e$ & $\beta$ & $\mu_0$ &
 $(b/a)_d$ & $r_h$& $L_B$ & $L_D$ & $L_T$ & $B/T$  & nuclear & error \\
 (1) & (2) & (3) & (4) & (5) & (6)
  & (7) & (8) & (9) & (10) & (11) & (12) & (13) & (14) \\ 
% & $[{\rm mag~arcsec^{-2}}]$&  & [arcsec] &  & $[{\rm mag~arcsec^{-2}}]$ &
%  & [arcsec] & [mag] & [mag] & [mag] & & [mag] & [mag] \\ 
\hline

 N 253 & 21.63 & 1.51 & 11.60 & 1/2 & 19.80 & 0.27 & 196.20 & 12.82 & 7.92 & 7.91 & 0.01 &  -  & 0.25 \\
 N 891 & - & - & - & - & - & - & - & - & - & - & - & - & no fit \\
 N 1068 & 17.96 & 0.63 & 9.98 & 1 & 18.83 & 0.91 & 30.90 & 10.78 & 9.48 & 9.10 & 0.21 & 11.83 & 0.15 \\
 N 2403 & 22.11 & 1.29 & 20.99 & 1 & 20.15 & 0.60 & 108.63 & 12.53 & 8.54 & 8.52 & 0.03 &  -  & 0.41 \\
 N 2841 & 18.79 & 0.69 & 10.96 & 1 & 19.64 & 0.41 & 68.41 & 11.29 & 9.43 & 9.23 & 0.15 & 13.33 & 0.25 \\
 N 2903 & 18.94 & 0.54 & 8.41 & 1 & 19.12 & 0.48 & 62.91 & 12.29 & 8.93 & 8.88 & 0.04 &  -  & 0.19 \\
 N 3031 & 18.46 & 0.74 & 25.76 & 1 & 19.07 & 0.61 & 147.00 & 9.04 & 6.85 & 6.71 & 0.12 & 12.94 & 0.15 \\
 N 3079 & 21.30 & 0.46 & 10.90 & 1 & 19.60 & 0.23 & 46.76 & 14.27 & 10.86 & 10.82 & 0.04 &  -  & 0.24 \\
 N 3198 & 21.43 & 0.87 & 5.75 & 3/4 & 20.41 & 0.38 & 64.08 & 14.95 & 10.44 & 10.43 & 0.02 &  -  & 0.18 \\
 N 3521 & 18.36 & 0.53 & 9.27 & 2/3 & 19.21 & 0.53 & 58.03 & 11.32 & 9.09 & 8.95 & 0.11 & 15.16 & 0.26 \\
 N 4192 & 17.91 & 0.75 & 3.02 & 1 & 19.99 & 0.23 & 82.57 & 13.12 & 10.02 & 9.96 & 0.06 &  -  & 0.20 \\
 N 4258 & 18.57 & 0.30 & 10.34 & 2/3 & 19.54 & 0.51 & 78.55 & 11.89 & 8.80 & 8.74 & 0.06 & 16.01 & 0.26 \\
 N 4321 & 19.12 & 0.81 & 7.96 & 1 & 20.49 & 0.70 & 85.91 & 12.15 & 9.21 & 9.14 & 0.06 &  -  & 0.19 \\
 N 4535 & 20.96 & 0.73 & 6.84 & 1 & 20.80 & 0.80 & 76.13 & 14.43 & 9.65 & 9.63 & 0.01 & 16.24 & 0.25 \\
 N 4536 & 19.46 & 0.72 & 4.10 & 1 & 21.09 & 0.67 & 52.52 & 14.06 & 10.94 & 10.88 & 0.05 &  -  & 0.27 \\
 N 4548 & 20.42 & 0.87 & 14.13 & 2/3 & 20.93 & 0.55 & 85.29 & 11.93 & 9.94 & 9.78 & 0.14 & 15.88 & 0.12 \\
 N 4565 & 22.18 & 0.94 & 55.23 & 1/2 & 20.50 & 0.07 & 199.38 & 10.52 & 10.11 & 9.54 & 0.41 &  -  & 0.37 \\
 N 4569 & 19.90 & 0.73 & 9.30 & 1 & 19.86 & 0.39 & 72.13 & 12.70 & 9.58 & 9.43 & 0.05 & 12.12 & 0.13 \\
 N 5194 & 19.13 & 1.21 & 11.98 & 1 & 20.01 & 0.73 & 112.25 & 10.84 & 8.13 & 8.04 & 0.08 & 13.98 & 0.27 \\
 N 5457 & 21.14 & 1.26 & 15.26 & 1 & 20.80 & 0.98 & 149.69 & 12.27 & 8.07 & 8.05 & 0.02 & 16.20 & 0.22 \\
 N 5907 & 21.41 & 0.29 & 21.23 & 3/4 & 20.46 & 0.10 & 124.91 & 13.27 & 10.56 & 10.48 & 0.08 &  -  & 0.18 \\
 N 6946 & 20.23 & 1.21 & 8.06 & 2/3 & 19.93 & 0.82 & 135.73 & 12.59 & 7.55 & 7.54 & 0.01 & 14.42 & 0.39 \\
 N 7331 & 18.66 & 0.38 & 16.65 & 1 & 19.68 & 0.42 & 64.47 & 10.90 & 9.57 & 9.29 & 0.23 & 14.55 & 0.26 \\

\hline
\multicolumn{14}{@{}l@{}}{\hbox to 0pt{\parbox{180mm}{\footnotesize
 \par\noindent
The colons denote unreliable data (see text).

(1) NGC Number. (2) Bulge effective luminosity $[{\rm
    mag~arcsec^{-2}}]$. (3) Bulge axial ratio. (4) Bulge effective
    radius [arcsec]. (5) Bulge shape index $\beta=1/n$. (6) Disk
    central luminosity $[{\rm mag~arcsec^{-2}}]$. (7) Disk axial
    ratio. (8) Disk scale length [arcsec]. (9) Total luminosity of
    bulge [mag]. (10) Total luminosity of disk [mag]. (11) Total
    luminosity of the galaxy [mag]. (12) Bulge-to-total (bulge plus
    disk) luminosity ratio. (13) Nucleus (central point source)
    luminosity [mag].  (14) Total residual in magnitude unit.  }\hss}}

% \footnotemark[(1)] Total luminosity of bulge. \footnotemark[(2)] Total luminosity of disk. \footnotemark[(3)] Bulge-to-total (bulge plus disk) luminosity ratio. }\hss}}  

\end{tabular}
 \end{center} 
% \label{tab.3}
\end{table*}

\begin{table*}
  \begin{center}
 \caption{\bf Results of $I$ band surface brightness fitting.}   
\begin{tabular}{c ccc c ccc ccc c c c}    
\hline\hline
 Name &  $\mu_e$ & $(b/a)_b$ & $r_e$ & $\beta$ & $\mu_0$ &
 $(b/a)_d$ & $r_h$& $L_B$ & $L_D$ & $L_T$ & $B/T$  & nuclear & error \\
 (1) & (2) & (3) & (4) & (5) & (6)
  & (7) & (8) & (9) & (10) & (11) & (12) & (13) & (14) \\ 
% & $[{\rm mag~arcsec^{-2}}]$&  & [arcsec] &  & $[{\rm mag~arcsec^{-2}}]$ &
%  & [arcsec] & [mag] & [mag] & [mag] & & [mag] & [mag] \\ 
\hline

 N 253 & 19.86 & 1.20 & 19.12 & 1/2 & 18.23 & 0.28 & 163.32 & 10.22 & 6.65 & 6.61 & 0.04 &  -  & 0.22 \\
 N 891 & - & - & - & - & - & - & - & - & - & - & - & - & no fit \\
 N 1068 & 16.79 & 0.59 & 9.77 & 1 & 17.59 & 0.92 & 25.29 & 9.72 & 8.67 & 8.23 & 0.25 & 11.04 & 0.13 \\
 N 2403 & 21.10 & 1.14 & 24.33 & 1 & 19.36 & 0.60 & 102.73 & 11.34 & 7.87 & 7.83 & 0.04 &  -  & 0.38 \\
 N 2841 & 17.52 & 0.72 & 10.46 & 1 & 18.35 & 0.42 & 63.81 & 10.09 & 8.26 & 8.06 & 0.15 & 12.64 & 0.19 \\
 N 2903 & 17.42 & 0.42 & 8.24 & 1 & 17.97 & 0.46 & 60.26 & 11.09 & 7.92 & 7.86 & 0.05 &  -  & 0.15 \\
 N 3031 & 16.99 & 0.76 & 17.99 & 1 & 17.39 & 0.62 & 98.66 & 8.32 & 5.95 & 5.83 & 0.10 & 12.18 & 0.10 \\
 N 3079 & 19.06 & 0.63 & 5.25 & 1 & 18.12 & 0.24 & 38.70 & 13.26 & 9.73 & 9.69 & 0.04 &  -  & 0.18 \\ 
 N 3198 & 20.17 & 0.90 & 6.97 & 3/4 & 19.33 & 0.37 & 59.28 & 13.24 & 9.56 & 9.52 & 0.03 &  -  & 0.15 \\
 N 3521 & 17.45 & 0.57 & 10.63 & 2/3 & 18.06 & 0.58 & 55.78 & 10.04 & 7.93 & 7.79 & 0.13 & 13.96 & 0.18 \\
 N 4192 & 16.55 & 0.62 & 3.36 & 1 & 18.80 & 0.23 & 74.53 & 11.73 & 9.03 & 8.94 & 0.08 &  -  & 0.18 \\
 N 4258 & 18.48 & 0.53 & 14.89 & 2/3 & 18.26 & 0.55 & 74.24 & 10.41 & 7.56 & 7.49 & 0.07 & 14.04 & 0.22 \\
 N 4321 & 18.09 & 0.73 & 8.61 & 1 & 19.42 & 0.75 & 87.64 & 11.06 & 8.03 & 7.97 & 0.06 &  -  & 0.08 \\
 N 4535 & 19.52 & 0.85 & 6.69 & 1 & 19.46 & 0.80 & 66.81 & 12.87 & 8.59 & 8.56 & 0.02 & 15.12 & 0.22 \\
 N 4536 & 17.62 & 0.64 & 4.22 & 1 & 19.62 & 0.66 & 43.20 & 12.29 & 9.90 & 9.79 & 0.10 &  -  & 0.26 \\
 N 4548 & 18.75 & 0.83 & 14.08 & 2/3 & 19.28 & 0.74 & 67.48 & 10.31 & 8.46 & 8.28 & 0.15 & 14.50 & 0.14 \\
 N 4565 & 19.98 & 0.79 & 37.57 & 1/2 & 18.91 & 0.08 & 160.37 & 9.33 & 8.78 & 8.27 & 0.38 &  -  & 0.33 \\
 N 4569 & 18.80 & 0.59 & 12.05 & 1 & 18.47 & 0.38 & 63.90 & 11.27 & 8.51 & 8.31 & 0.07 & 10.79 & 0.13 \\
 N 4736 & 16.70 & 0.99 & 17.66 & 2/3 & 18.42 & 0.70 & 75.92 & 7.58 & 7.42 & 6.75 & 0.46 &  -  & 0.11\\
 N 5194 & 18.03 & 1.24 & 11.86 & 1 & 18.94 & 0.75 & 101.70 & 9.73 & 7.24 & 7.13 & 0.09 & 13.27 & 0.26 \\
 N 5457 & 20.00 & 1.31 & 14.42 & 1 & 19.67 & 1.00 & 120.95 & 11.22 & 7.33 & 7.30 & 0.03 & 15.49 & 0.18 \\
 N 5907 & 19.75 & 0.36 & 14.66 & 3/4 & 18.75 & 0.10 & 94.23 & 12.19 & 9.40 & 9.32 & 0.07 &  -  & 0.19 \\
 N 6946 & 19.71 & 0.82 & 19.07 & 2/3 & 18.98 & 0.82 & 126.33 & 10.63 & 6.75 & 6.72 & 0.03 & 12.29 & 0.28 \\
 N 7331 & 17.48 & 0.41 & 15.78 & 1 & 18.15 & 0.46 & 51.51 & 9.77 & 8.44 & 8.15 & 0.22 & 13.28 & 0.21 \\

\hline
\multicolumn{14}{@{}l@{}}{\hbox to 0pt{\parbox{180mm}{\footnotesize
 \par\noindent
The colons denote unreliable data (see text).

(1) NGC Number. (2) Bulge effective luminosity $[{\rm
    mag~arcsec^{-2}}]$. (3) Bulge axial ratio. (4) Bulge effective
    radius [arcsec]. (5) Bulge shape index $\beta=1/n$. (6) Disk
    central luminosity $[{\rm mag~arcsec^{-2}}]$. (7) Disk axial
    ratio. (8) Disk scale length [arcsec]. (9) Total luminosity of
    bulge [mag]. (10) Total luminosity of disk [mag]. (11) Total
    luminosity of the galaxy [mag]. (12) Bulge-to-total (bulge plus
    disk) luminosity ratio. (13) Nucleus (central point source)
    luminosity [mag].  (14) Total residual in magnitude unit.  }\hss}}  

\end{tabular}
  \end{center}  
%  \label{tab.4}
\end{table*}

\begin{table*}[t]
\begin{center}
\caption{\bf Results of $J$ band surface brightness fitting.}
\begin{tabular}{c ccc c ccc ccc c c c}    
\hline\hline
 Name &  $\mu_e$ & $(b/a)_b$ & $r_e$ & $\beta$ & $\mu_0$ &
 $(b/a)_d$ & $r_h$& $L_B$ & $L_D$ & $L_T$ & $B/T$  & nuclear & error \\
 (1) & (2) & (3) & (4) & (5) & (6)
  & (7) & (8) & (9) & (10) & (11) & (12) & (13) & (14) \\ 
% & $[{\rm mag~arcsec^{-2}}]$&  & [arcsec] &  & $[{\rm mag~arcsec^{-2}}]$ &
%  & [arcsec] & [mag] & [mag] & [mag] & & [mag] & [mag] \\ 
\hline

 N253 & 16.50 & 0.57 & 12.74 & 1/2 & 16.39 & 0.45 & 133.50 & 8.53 & 4.67 & 4.64 & 0.03 &  -  & 0.11 \\
 N 891 & - & - & - & - & - & - & - & - & - & - & - & - & no fit \\
 N2841 & 16.61 & 0.40 & 10.87 & 1 & 16.92 & 0.52 & 56.15 & 9.71 & 6.88 & 6.76 & 0.07 & 10.32 & 0.13 \\
 N2903 & 14.94: & 0.38: & 5.18: & 1: & 17.17: & 0.28: & 122.86: & 9.70: & 6.13: & 6.09: & 0.04: &  -  & 0.10 \\
 N3031 & 15.86 & 0.65 & 26.19 & 1 & 16.29 & 0.57 & 98.34 & 6.54 & 4.95 & 4.70 & 0.18 & 8.84 & 0.11 \\
 N3079 & 15.80 & 0.59 & 3.93 & 1 & 16.98 & 0.13 & 54.70 & 10.70 & 8.48 & 8.35 & 0.11 &  -  & 0.14 \\
 N3198 & 17.76: & 0.82: & 1.08: & 3/4: & 17.68: & 0.36: & 32.72: & 14.54: & 9.23: & 9.22: & 0.01: &  -  & 0.17 \\
 N3521 & 16.17 & 0.74 & 10.27 & 2/3 & 16.45 & 0.53 & 49.58 & 8.55 & 6.67 & 6.49 & 0.15 & - & 0.09 \\
 N3628 & 19.09 & 0.86 & 16.87 & 1/2 & 16.84 & 0.20 & 51.14 & 10.08 & 8.05 & 7.89 & 0.13 &  -  & 0.16 \\
 N4258 & 16.43: & 0.61: & 7.31: & 2/3: & 16.37: & 0.54: & 38.95: & 9.75: & 7.10: & 7.00: & 0.08: & 12.76: & 0.15 \\
 N4303 & 15.73 & 0.99 & 2.27 & 1 & 16.71 & 0.75 & 22.38 & 11.26 & 8.28 & 8.20 & 0.06 & 13.11 & 0.18 \\
 N4321 & 16.72: & 0.69: & 7.49: & 1: & 18.25: & 0.67: & 85.94: & 10.06: & 7.03: & 6.96: & 0.06: &  -  & 0.14 \\
 N4535 & 18.58: & 0.69: & 6.89: & 1: & 18.77: & 0.91: & 71.38: & 12.10: & 7.61: & 7.59: & 0.02: & 12.98: & 0.13 \\
 N4536 & - & - & - & - & - & - & - & - & - & - & - & - & no fit \\
 N4631 & 17.08 & 0.15 & 16.71 & 1 & 17.75 & 0.19 & 64.16 & 10.31 & 8.55 & 8.35 & 0.17 &  -  & 0.27 \\
 N4736 & 13.49: & 1.08: & 2.29: & 2/3: & 15.28: & 0.76: & 23.35: & 8.63: & 6.74: & 6.56: & 0.15: &  -  & 0.24 \\
 N5907 & 17.90 & 0.49 & 8.91 & 3/4 & 16.89 & 0.09 & 79.96 & 11.09 & 8.01 & 7.95 & 0.06 &  -  & 0.12 \\
 N6946 & 17.62 & 0.86 & 10.92 & 2/3 & 17.85 & 0.75 & 110.02 & 9.70 & 5.99 & 5.95 & 0.03 & 11.92 & 0.37 \\
 N7331 & 16.20 & 0.38 & 17.62 & 1 & 16.87 & 0.47 & 55.31 & 8.32 & 6.98 & 6.69 & 0.22 & 11.15 & 0.28 \\

\hline
\multicolumn{14}{@{}l@{}}{\hbox to 0pt{\parbox{180mm}{\footnotesize
 \par\noindent
The colons denote unreliable data (see text).

(1) NGC Number. (2) Bulge effective luminosity $[{\rm
    mag~arcsec^{-2}}]$. (3) Bulge axial ratio. (4) Bulge effective
    radius [arcsec]. (5) Bulge shape index $\beta=1/n$. (6) Disk
    central luminosity $[{\rm mag~arcsec^{-2}}]$. (7) Disk axial
    ratio. (8) Disk scale length [arcsec]. (9) Total luminosity of
    bulge [mag]. (10) Total luminosity of disk [mag]. (11) Total
    luminosity of the galaxy [mag]. (12) Bulge-to-total (bulge plus
    disk) luminosity ratio. (13) Nucleus (central point source)
    luminosity [mag].  (14) Total residual in magnitude unit.   }\hss}}  

\end{tabular}
\end{center}
% \label{tab.5}
\end{table*} 

\begin{table*}[t]
 \begin{center}
\caption{\bf Colors of bulges and disks.}  
\begin{tabular}{ccccc}
\hline\hline  
 Name & $(V-I)_b$ & $(V-I)_d$ & $(I-J)_b$ & $(I-J)_d$ \\
 (1) & (2) & (3) & (4) & (5) \\

\hline

 N 253 & 2.60 & 1.27 & 1.69 & 1.98  \\ 
 N 891 & - & - & - & - \\ 
 N 1068 & 1.06 & 0.82 & - & -  \\
 N 2403 & 1.19 & 0.67 & - & -  \\
 N 2841 & 1.20 & 1.17 & 0.37 & 1.38  \\
 N 2903 & 1.19 & 1.01 & 1.39: & 1.79:  \\
 N 3031 & 0.72 & 0.90 & 1.79 & 1.00  \\
 N 3079 & 1.01 & 1.13 & 2.56 & 1.25   \\
 N 3198 & 1.71 & 0.89 & -1.31: & 0.33:  \\
 N 3521 & 1.29 & 1.15 & 1.49 & 1.27  \\
 N 4192 & 1.39 & 1.00 & - & -  \\
 N 4258 & 1.48 & 1.24 & 0.66: & 0.46:  \\
 N 4321 & 1.09 & 1.18 & 1.01: & 1.00:  \\
 N 4535 & 1.56 & 1.06 & 0.77: & 0.97:  \\
 N 4536 & 1.77 & 1.03 & - & -  \\
 N 4548 & 1.61 & 1.48 & - & -  \\
 N 4565 & 1.18 & 1.32 & - & -  \\
 N 4569 & 1.43 & 1.08 & - & -  \\
 N 4736 & - & - & -1.04: & 0.68:  \\
 N 5194 & 1.10 & 0.89 & - & -  \\
 N 5457 & 1.05 & 0.74 & - & -  \\
 N 5907 & 1.08 & 1.17 & 1.10 & 1.39  \\
 N 6946 & 1.96 & 0.80 & 0.93 & 0.77  \\
 N 7331 & 1.13 & 1.14 & 1.45 & 1.46  \\

\hline

\multicolumn{5}{@{}l@{}}{\hbox to 0pt{\parbox{180mm}{\footnotesize
 \par\noindent
The colons denote unreliable data (see text).

(1) NGC Number. (2) Bulge $V-I$. 
(3) Disk $V-I$. (4) Bulge $I-J$. 
(5) Disk $I-J$. }\hss}}  

\end{tabular}
 \end{center} 
% \label{tab.6}
\end{table*}

\begin{table*}
\begin{center}
\caption{\bf $V$ band $M/L$ ratios and halo parameters.}
\begin{tabular}{cccccccc}    
\hline\hline   
 Name & $(M/L_V)_b$ & $\sigma_{(M/L_V)b}$ & $(M/L_V)_d$ & $\sigma_{(M/L_V)d}$ & $\sigma_h$ & $R_h$  & rms \\ 
  (1)    &  (2)  &  (3) &  (4)  & (5) & (6)  & (7) & (8) \\ 
%      &             &             &   $10^2[{\rm km~s^{-1}}]$  &  [kpc] &  \\ 
\hline

 N 253 & 104.94: & 85.88 & 3.17 & 0.59 & - & - & 26.07 \\
 N 891 & - & - & - & - & - & - & no fit \\
 N 1068 & 1.46 & 0.46 & 1.29 & 0.43 & - & - & 25.66 \\
 N 2403 & 17.10: & 7.93 & 1.54 & 0.73 & 110.74 & 0.31 & 3.03 \\
 N 2841 & 2.06 & 0.40 & 4.04 & 0.61 & - & - & 47.91 \\
 N 2903 & 4.21 & 1.26 & 2.22 & 0.48 & 157.45 & 0.27 & 11.77 \\
 N 3031 & 4.89 & 1.02 & 1.36 & 0.44 & 100.44 & 0.20 & 21.71 \\
 N 3079 & 54.28: & 24.20 & 1.30 & 0.33 & 114.16 & 0.09 & 35.63 \\
 N 3198 & 2.21: & 1.64 & 1.71 & 0.57 & 125.95 & 0.21 & 5.19 \\
 N 3521 & 2.55 & 0.54 & 2.34 & 0.66 & 138.95 & 0.25 & 14.72 \\
 N 4192 & 3.51 & 0.64 & 1.63 & 0.31 & - & - & 33.38 \\
 N 4258 & 6.89 & 1.27 & 1.74 & 0.50 & 202.97 & 0.51 & 31.09 \\
 N 4321 & 4.91 & 1.63 & 3.41 & 1.29 & - & - & 4.11 \\
 N 4535 & 18.70: & 9.93 & 2.26: & 0.71 & - & - & 12.24 \\
 N 4536 & 7.42 & 1.89 & 2.61: & 1.21 & - & - & 20.46 \\
 N 4548 & 6.94 & 2.48 & 1.04: & 0.52 & - & - & 13.56 \\
 N 4565 & 22.33: & 8.54 & 1.77: & 0.63 & - & - & 17.82 \\
 N 4569 & 1.55 & 0.94 & 1.76 & 0.40 & - & - & 24.83 \\
 N 5194 & 8.79 & 2.33 & 2.45 & 0.57 & - & - & 46.43 \\
 N 5457 & 49.88: & 19.06 & 3.24: & 2.07 & - & - & 3.67 \\
 N 5907 & 34.32: & 13.39 & 2.49 & 0.47 & - & - & 19.66 \\
 N 6946 & 23.46: & 12.25 & 2.43 & 1.02 & - & - & 3.70 \\
 N 7331 & 2.86 & 0.64 & 1.38 & 0.41 & - & - & 16.58 \\

\hline

\multicolumn{6}{@{}l@{}}{\hbox to 0pt{\parbox{180mm}{\footnotesize
 \par\noindent
The colons denote unreliable data (see text).

 (1) NGC Number. (2) Bulge $M/L$ at the $V$ band.  (3) 1$\sigma$ error
of the bulge $M/L$.  (4) Disk $M/L$ at the $V$ band.  \\ (5) 1$\sigma$
error of the disk $M/L$.  (6) Halo velocity dispersion in $10^2[{\rm
km~s^{-1}}]$ unit. (7) Halo core radius [kpc].  \\ (8) Residual in
$[{\rm km~s^{-1}}]$ unit. }\hss}}

\end{tabular}
\end{center}
% \label{tab.7}
\end{table*} 

\begin{table*}
\begin{center}
\caption{\bf $I$ band $M/L$ ratios and halo parameters.}
\begin{tabular}{cccccccc}    
\hline\hline   
 Name & $(M/L_I)_b$ & $\sigma_{(M/L_I)b}$ & $(M/L_I)_d$ & $\sigma_{(M/L_I)d}$  & $\sigma_h$ & $R_h$  & rms \\
  (1)    &  (2)  &  (3) &  (4)  & (5) & (6)  & (7) & (8) \\ 
%      &             &             & $10^2[{\rm km~s^{-1}}]$    &  [kpc] &  \\ 
\hline

 N 253 & 21.04: & 7.39 & 1.60 & 0.33 & - & - & 18.06 \\
 N 891 & - & - & - & - & - & - & no fit \\
 N 1068 & 1.11 & 0.32 & 0.74 & 0.31 & - & - & 22.42 \\
 N 2403 & 14.00: & 6.22 & 1.42 & 0.72 & 109.64 & 0.28 & 3.13 \\
 N 2841 & 1.08 & 0.21 & 2.59 & 0.41 & 745.46 & 1.93 & 43.67 \\
 N 2903 & 3.53 & 1.10 & 1.48 & 0.35 & 151.97 & 0.23 & 11.16 \\
 N 3031 & 3.14 & 0.53 & 0.97 & 0.22 & 118.65 & 0.15 & 29.76 \\
 N 3079 & 16.40: & 5.02 & 0.90 & 0.25 & 128.75 & 0.09 & 20.59 \\
 N 3198 & 0.23: & 0.81 & 1.34 & 0.44 & 123.77 & 0.17 & 5.67 \\
 N 3521 & 1.69 & 0.36 & 1.69 & 0.46 & 135.61 & 0.21 & 13.64 \\
 N 4192 & 1.87 & 0.31 & 1.10 & 0.26 & - & - & 32.20 \\
 N 4258 & 6.29 & 1.65 & 0.97 & 0.30 & 187.18 & 0.39 & 24.53 \\
 N 4321 & 4.16 & 1.50 & 2.46 & 0.91 & - & - & 4.02 \\
 N 4535 & 5.81: & 2.83 & 1.48 & 0.31 & - & - & 12.67 \\
 N 4536 & 3.34 & 0.90 & 1.80: & 0.77 & - & - & 14.09 \\
 N 4548 & 2.27 & 0.86 & 0.75 & 0.13 & - & - & 13.18 \\
 N 4565 & 6.82: & 1.61 & 1.14 & 0.23 & 84.42 & 0.40 & 13.84 \\
 N 4569 & 0.77 & 0.51 & 0.97 & 0.23 & - & - & 24.08 \\
 N 4736 & 0.89 & 0.28 & 1.23 & 0.72 & 63.13 & 0.09 & 4.41 \\
 N 5194 & 5.68 & 1.40 & 2.14 & 0.50 & - & - & 48.42 \\
 N 5457 & 17.73: & 8.61 & 3.29: & 1.95 & - & - & 2.79 \\
 N 5907 & 12.62: & 4.28 & 1.28 & 0.22 & 120.56 & 0.30 & 12.39 \\
 N 6946 & 17.09: & 21.16 & 1.90 & 0.87 & 196.40 & 0.69 & 6.84 \\
 N 7331 & 1.67 & 0.35 & 0.91 & 0.26 & 531.05 & 0.83 & 14.05 \\

\hline

\multicolumn{6}{@{}l@{}}{\hbox to 0pt{\parbox{180mm}{\footnotesize
 \par\noindent
The colons denote unreliable data (see text).

(1) NGC Number. (2) Bulge $M/L$ at the $I$ band.  (3) 1$\sigma$ error
of the bulge $M/L$.  (4) Disk $M/L$ at the $I$ band.  \\ (5) 1$\sigma$
error of the disk $M/L$.  (6) Halo velocity dispersion in $10^2[{\rm
km~s^{-1}}]$ unit. (7) Halo core radius [kpc].  \\(8) Residual in
$[{\rm km~s^{-1}}]$ unit. }\hss}}

\end{tabular}
\end{center}
% \label{tab.8}
\end{table*}

\begin{table*}
\begin{center}
\caption{\bf $J$ band $M/L$ ratios and halo parameters.}
\begin{tabular}{cccccccc}    
\hline\hline   
 Name & $(M/L_J)_b$ & $\sigma_{(M/L_J)b}$ & $(M/L_J)_d$ & $\sigma_{(M/L_J)d}$ & $\sigma_h$ & $R_h$ & rms \\
  (1)    &  (2)  &  (3) &  (4)  & (5) & (6)  & (7) & (8) \\ 
%      &       &      &     &        & $10^2[{\rm km~s^{-1}}]$   &  [kpc] &   \\
\hline

 N 253 & 2.42: & 0.56 & 0.61 & 0.10 & - & - & 21.35 \\
 N 891 & - & - & - & - & - & - & no fit \\
 N 1808 & - & - & - & - & - & - & no fit \\
 N 2841 & 1.01 & 0.28 & 1.18 & 0.27 & 347.87 & 0.71 & 32.75 \\
 N 2903 & 0.86: & 0.14 & 0.62: & 0.12 & - & - & 23.69 \\
 N 3031 & 1.33 & 0.23 & 0.39 & 0.18 & 121.84 & 0.11 & 26.67 \\
 N 3079 & 1.61: & 0.26 & 0.40 & 0.08 & 31.77 & 0.00 & 16.72 \\
 N 3198 & 0.25: & 0.48 & 0.51: & 0.18 & 119.37: & 0.05: & 8.62 \\
 N 3521 & 0.87 & 0.20 & 0.62 & 0.18 & 131.52 & 0.13 & 13.14 \\
 N 3628 & 6.81: & 2.16 & 0.61 & 0.15 & 191.25 & 0.19 & 22.39 \\
 N 4258 & 0.52: & 0.10 & 0.95: & 0.24 & 178.60: & 0.29: & 27.18 \\
 N 4303 & 0.98 & 0.43 & 0.47 & 0.29 & 135.24 & 0.11 & 1.33 \\
 N 4321 & 1.35: & 0.41 & 1.41: & 0.42 & - & - & 4.70 \\
 N 4535 & 7.28: & 3.05 & 1.33: & 0.25 & - & - & 12.71 \\
 N 4536 & - & - & - & - & - & - & no fit \\
 N 4631 & 0.55: & 0.42 & 1.27 & 0.28 & 120.58 & 0.22 & 13.16 \\
 N 4736 & 0.15: & 0.05 & 0.78: & 0.22 & 110.58: & 0.07: & 9.19 \\
 N 5907 & 3.43: & 0.68 & 0.43 & 0.07 & 149.43 & 0.25 & 17.47 \\
 N 6946 & 5.15: & 1.82 & 1.22 & 0.53 & 155.43 & 0.36 & 4.15 \\
 N 7331 & 0.79 & 0.20 & 0.37 & 0.14 & 665.95 & 1.09 & 13.47 \\

\hline

\multicolumn{6}{@{}l@{}}{\hbox to 0pt{\parbox{180mm}{\footnotesize
 \par\noindent
The colons denote unreliable data (see text).

(1) NGC Number. (2) Bulge $M/L$ at the $J$ band.  (3) 1$\sigma$ error
of the bulge $M/L$.  (4) Disk $M/L$ at the $J$ band.  \\ (5) 1$\sigma$
error of the disk $M/L$.  (6) Halo velocity dispersion in $10^2[{\rm
km~s^{-1}}]$ unit. (7) Halo core radius [kpc].  \\ (8) Residual in
$[{\rm km~s^{-1}}]$ unit. }\hss}}

\end{tabular}
\end{center}
% \label{tab.9}
\end{table*} 

\begin{figure*}
  \begin{center}
    \FigureFile(180mm,180mm){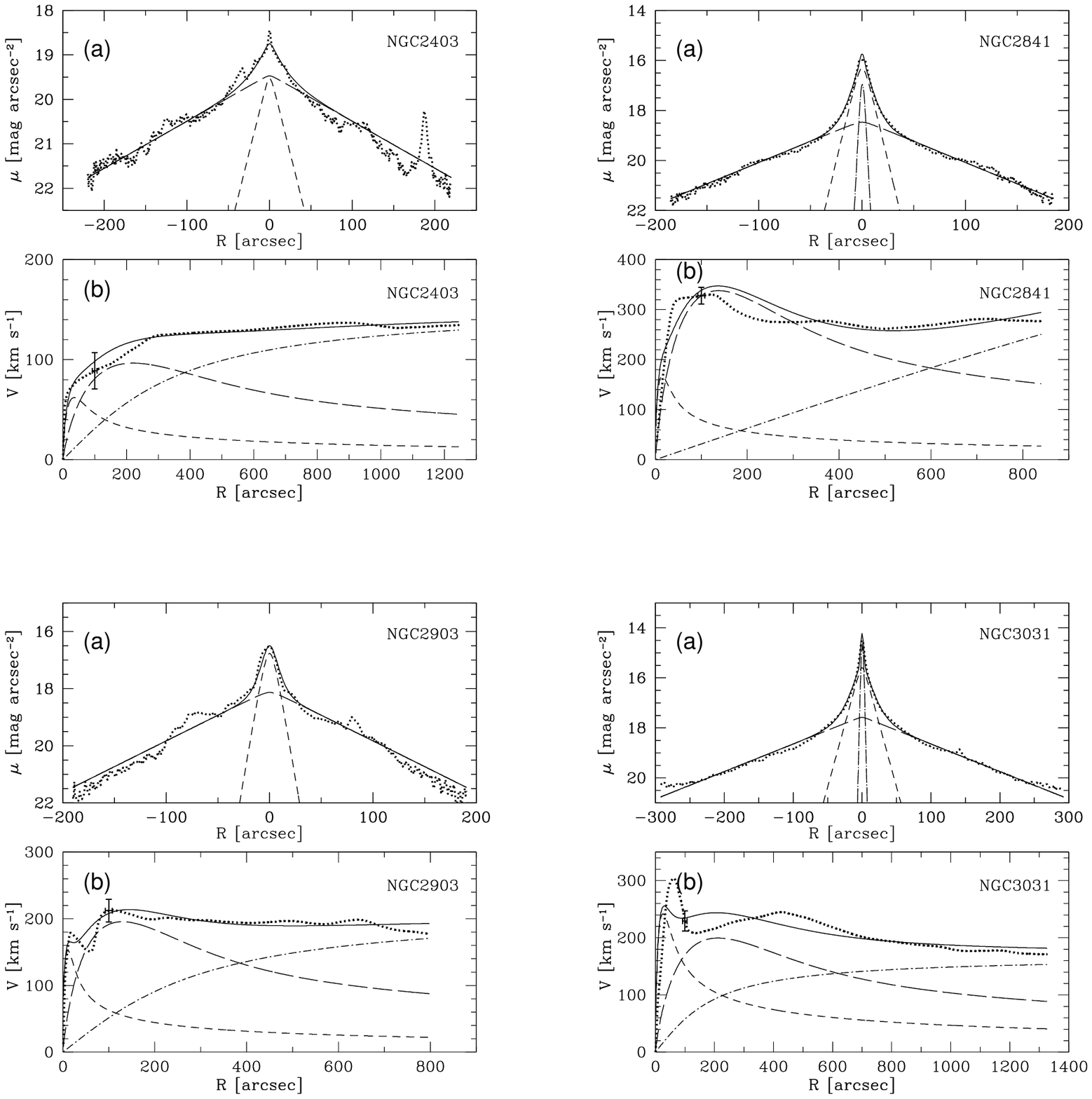}
  \end{center}

\end{figure*}

\begin{figure*}
  \begin{center}
    \FigureFile(180mm,180mm){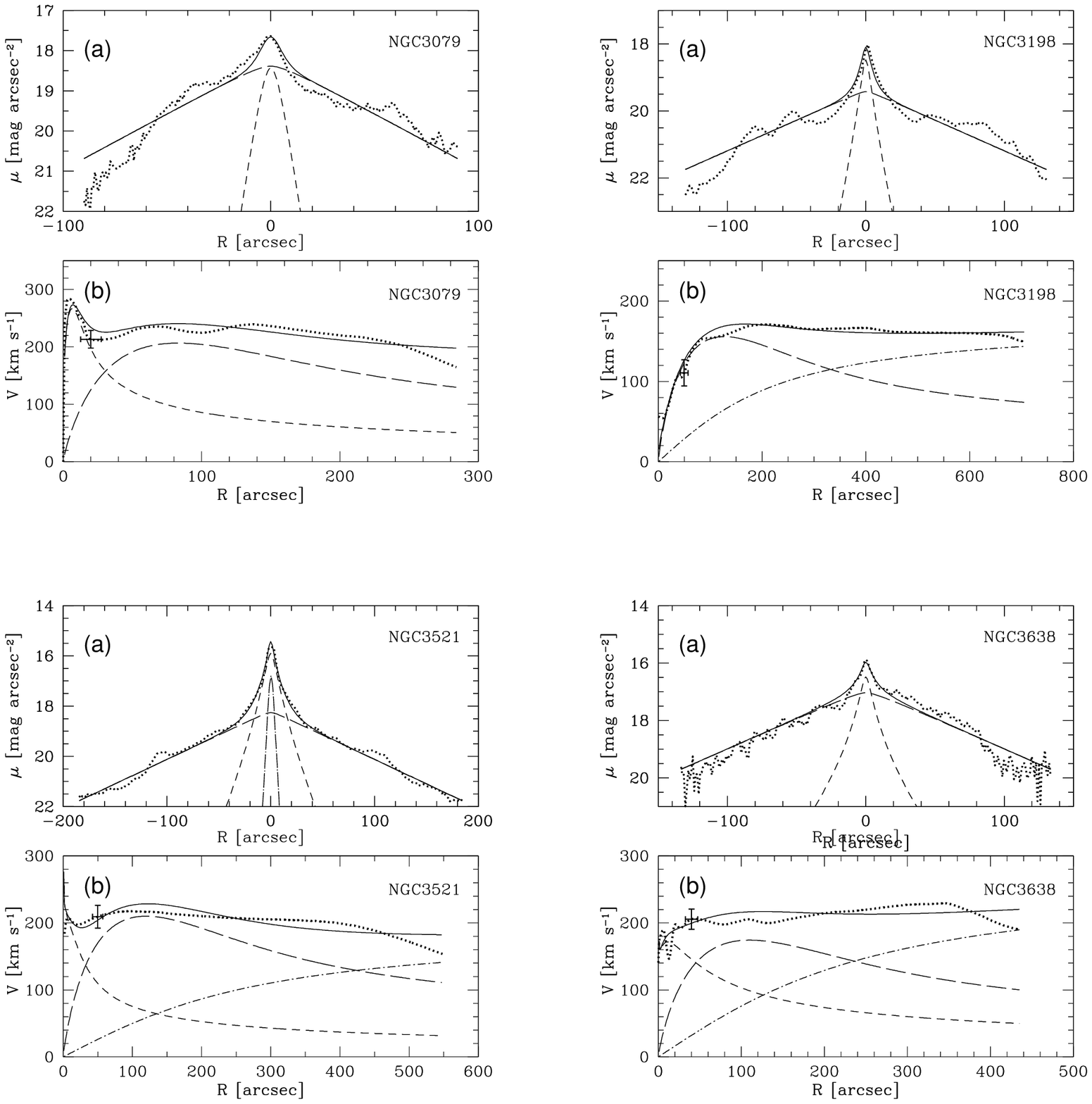}
  \end{center}

\end{figure*}

\begin{figure*}
  \begin{center}
    \FigureFile(180mm,180mm){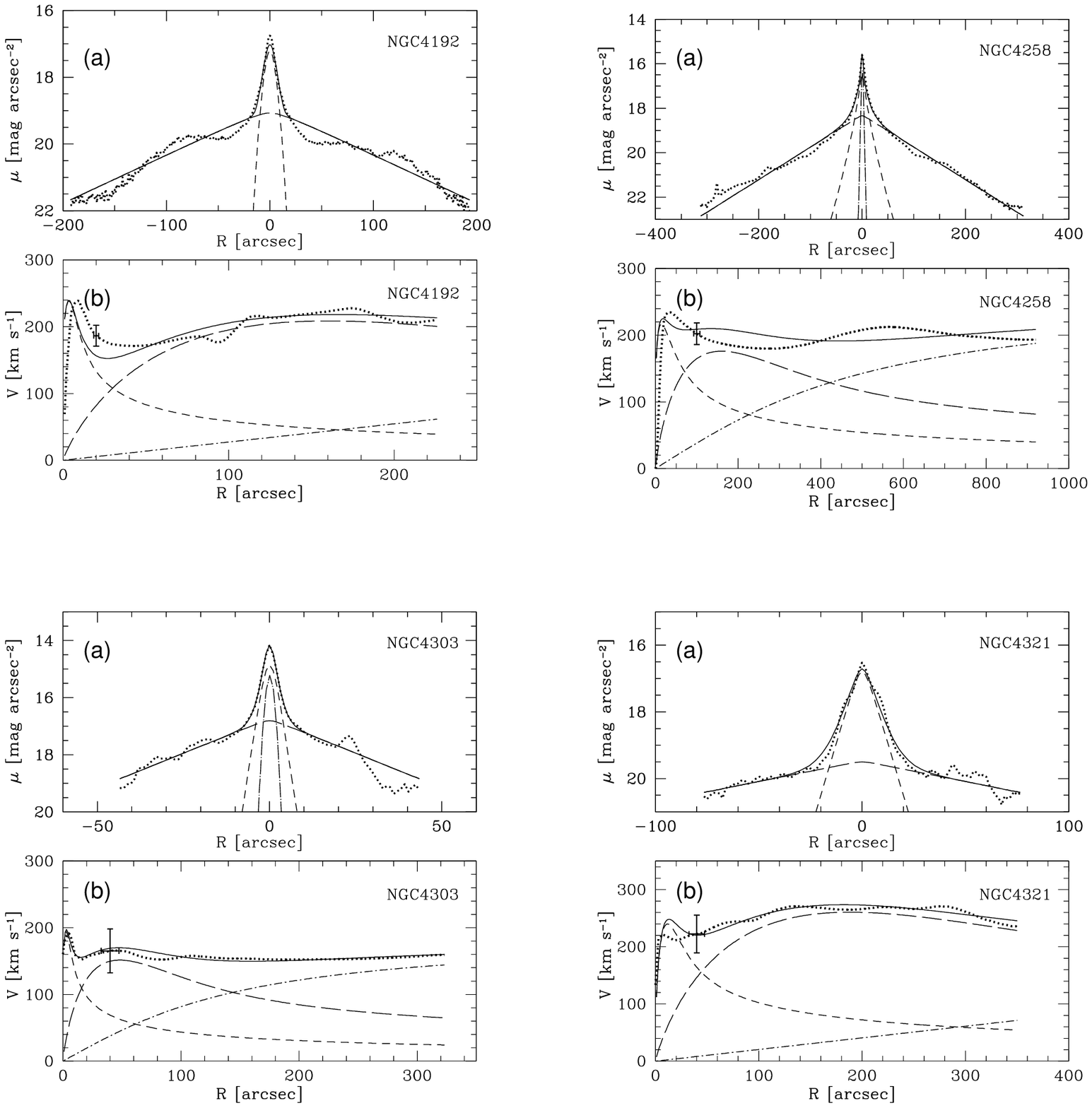}
  \end{center}

\end{figure*}

\begin{figure*}
  \begin{center}
    \FigureFile(180mm,180mm){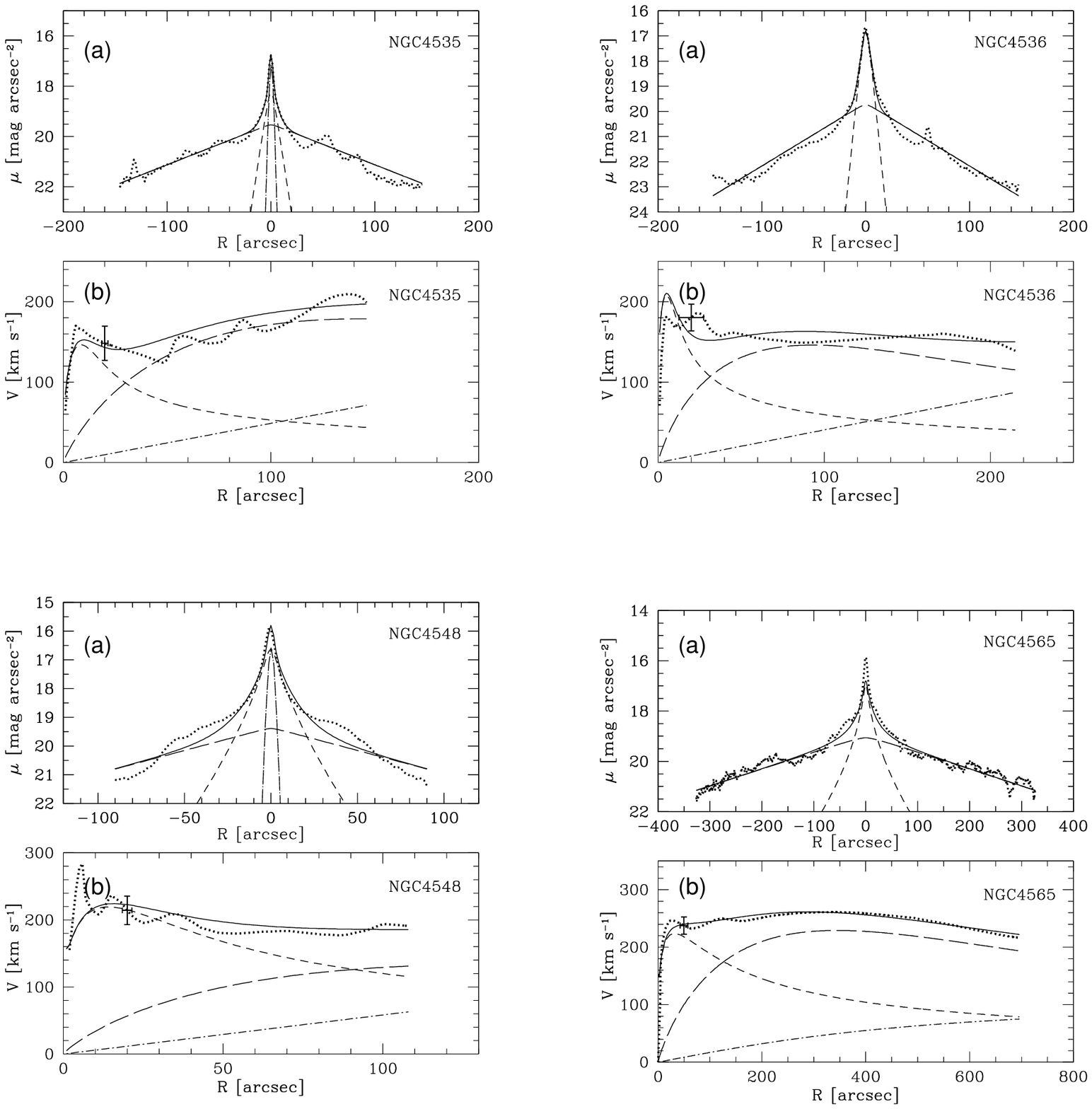}
  \end{center}

\end{figure*}

\begin{figure*}
  \begin{center}
    \FigureFile(180mm,180mm){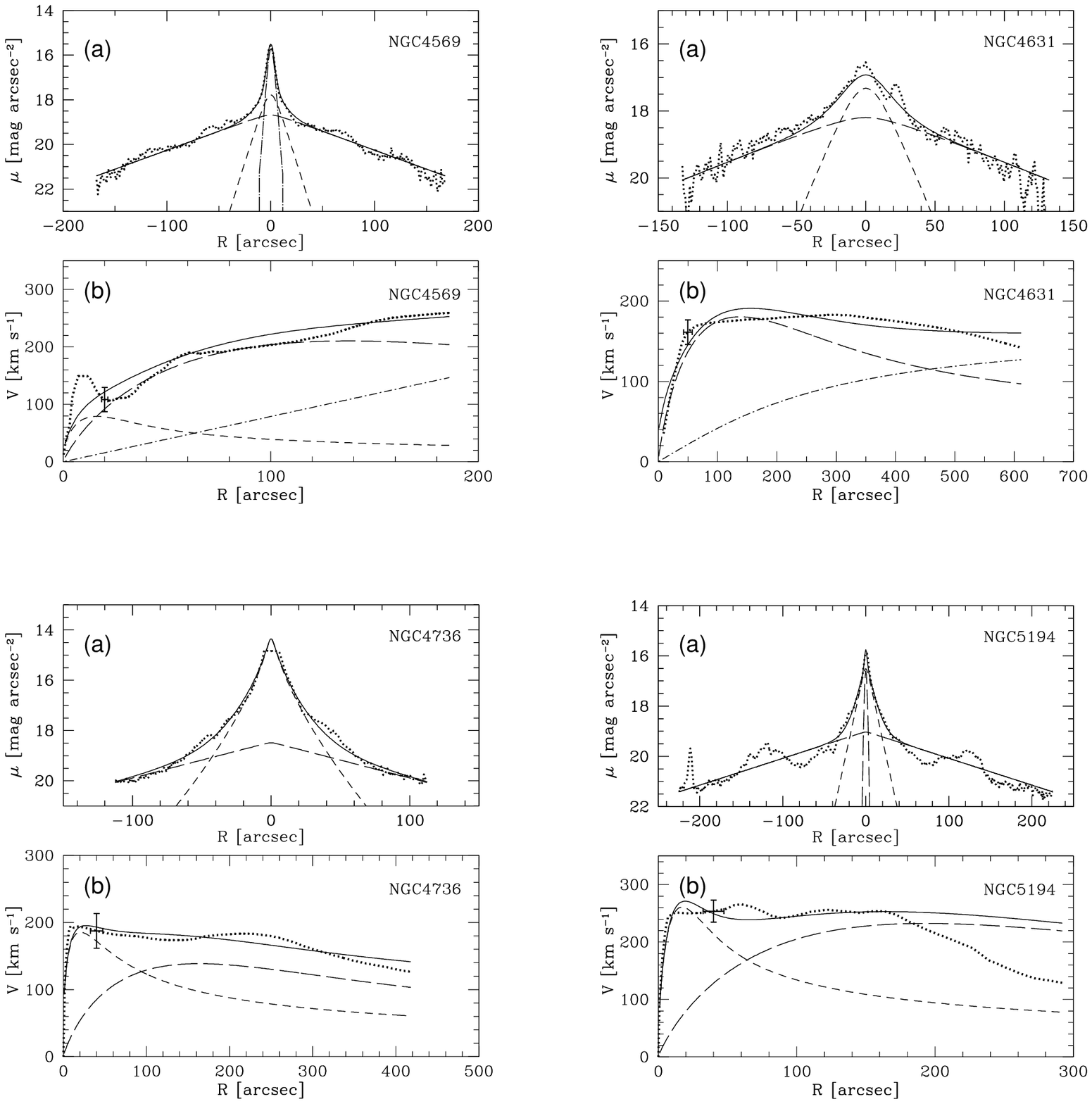}
  \end{center}

\end{figure*}

\begin{figure*}
  \begin{center}
    \FigureFile(180mm,180mm){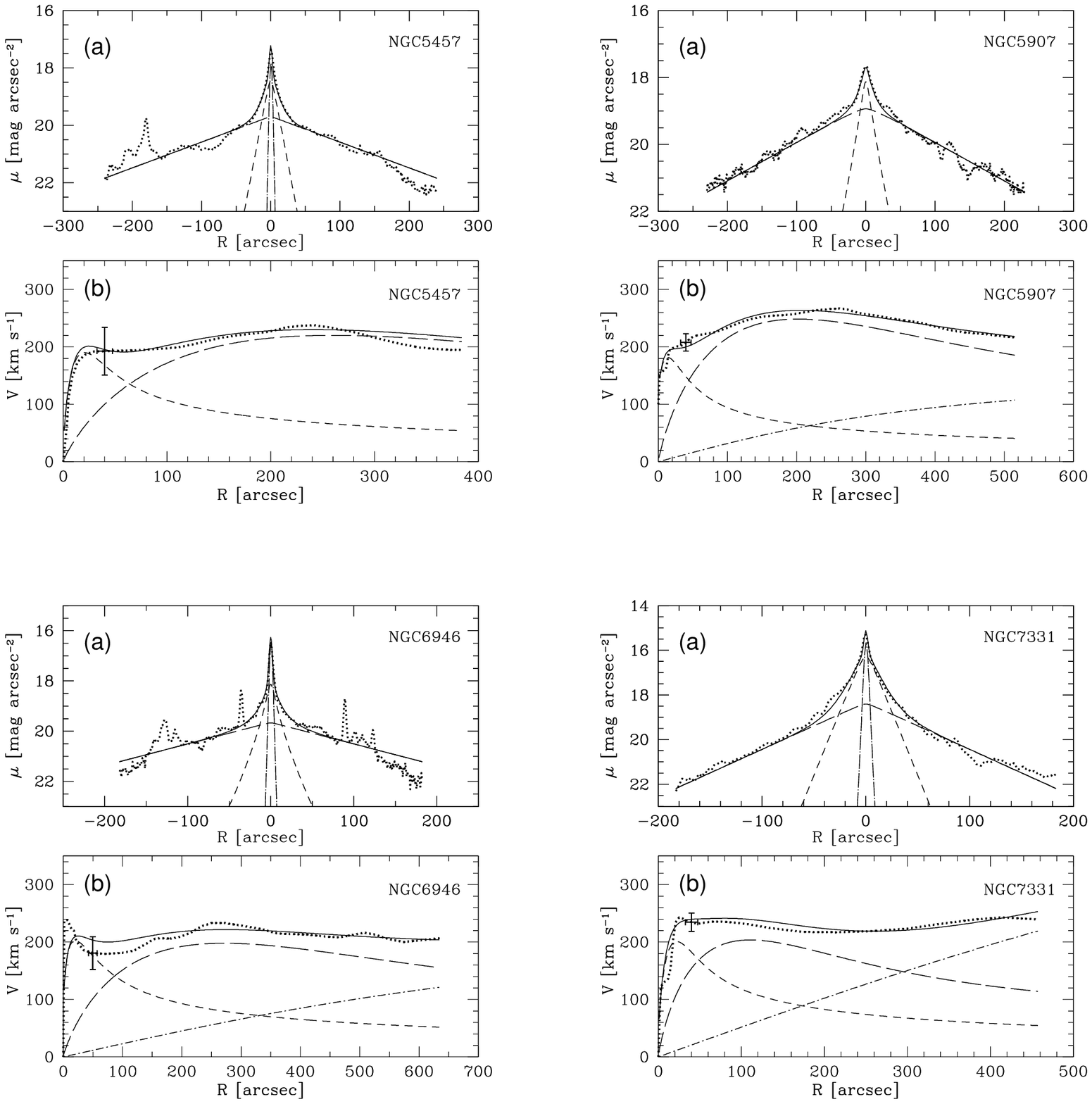}
  \end{center}

\end{figure*}

\begin{figure}
  \begin{center} \FigureFile(80mm,80mm){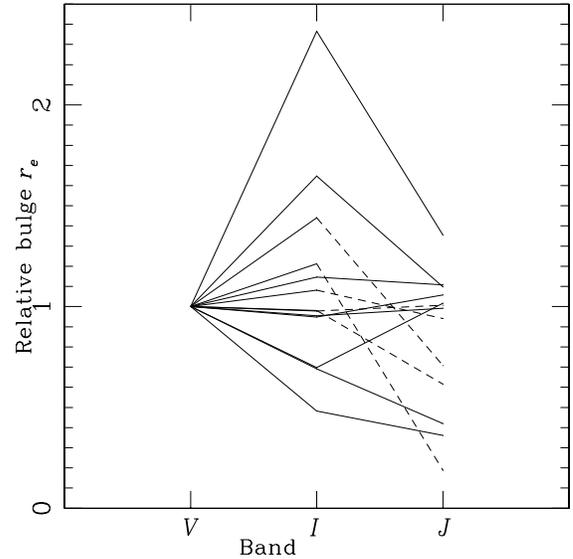} \end{center}
    \caption{The trend of bulge effective radius $r_e$ on bands for 13
    galaxies. The $r_e$ is normalized to that of $V$ band. The dashed
    lines are unreliable data (NGC 2903, 3198, 4258, 4321 and 4535 in
    the $J$ band).}
\label{fig.7}
\end{figure}

\begin{figure}
  \begin{center} \FigureFile(80mm,80mm){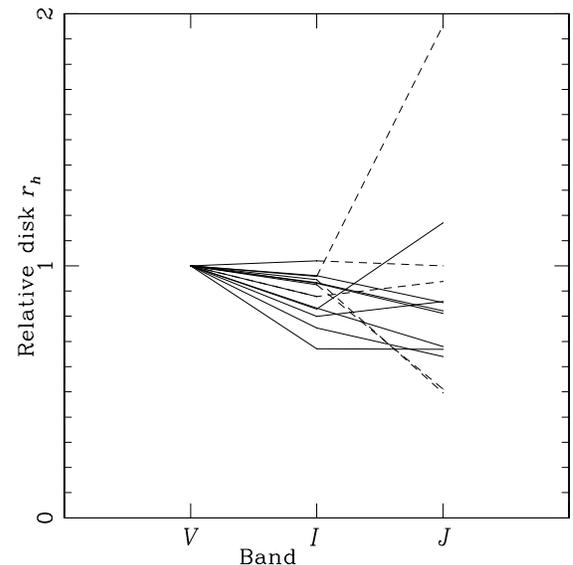} \end{center}
    \caption{The trend of disk scale length $r_h$ on bands for 13
    galaxies. The $r_h$ is normalized to that of $V$ band. The dashed
    lines are unreliable data (NGC 2903, 3198, 4258, 4321 and 4535 in
    the $J$ band).}
\label{fig.8}
\end{figure}

\begin{figure}
  \begin{center} \FigureFile(80mm,80mm){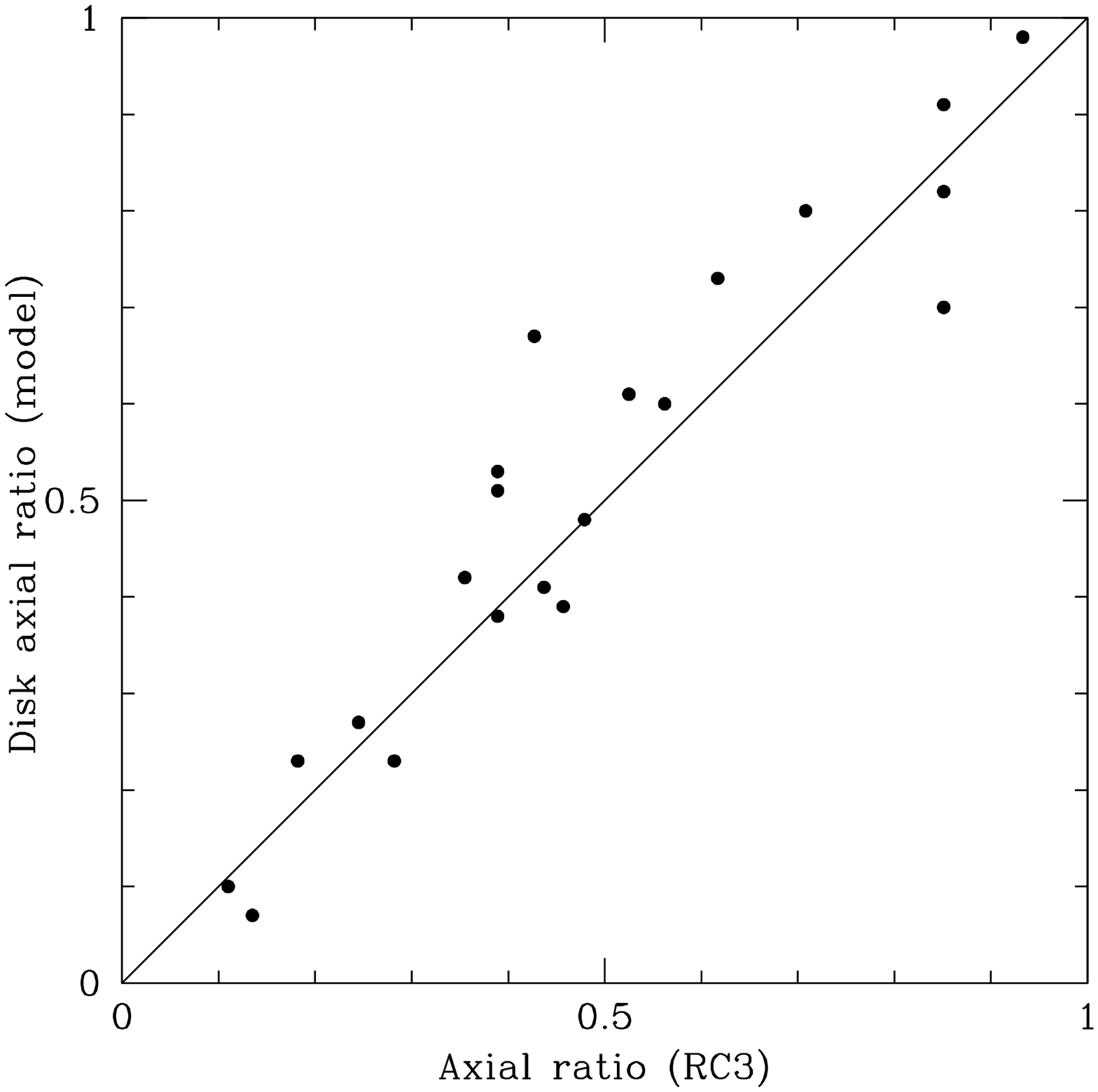} \end{center}
    \caption{The comparison of the disk axial ratio $(b/a)_d$ obtained
    from the fitting in the present study with the catalogue axial ratios
    taken from RC3. The model $(b/a)_d$ is for $V$ band, while that of
    RC3 is derived from the contour of 25 $B$ magnitude. The solid
    line shows the equality.}
\label{fig.9}
\end{figure}

\subsection{Dust Extinction}

\begin{figure}
  \begin{center} \FigureFile(80mm,80mm){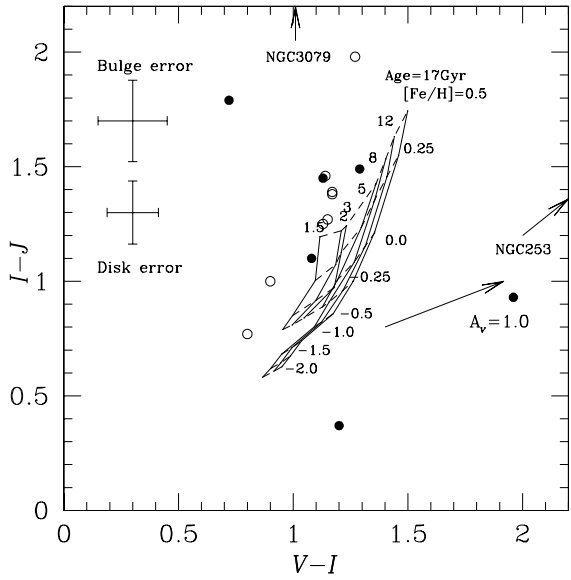} \end{center}
    \caption{Color-color diagram for the bulge (filled circles) and
    disks (open circles). The error bars represent the typical errors
    for bulge and disk. The models of Worthey (1994) of the same age
    are connected by solid lines, while the models of the same
    metallicity are connected by dashed lines. The bulges of NGC 253
    and NGC 3079 are outside the figure and locate in the directions
    of the arrows.}  \label{fig.10}
\end{figure}

The bulge and disk colors are listed in table 6. Figure \ref{fig.10}
shows the color-color ($V-I$ versus $I-J$) diagram. The mesh in the
figure represents W94 model of single starburst with Salpeter IMF,
the age = 1.5 to 17 Gyr, and [Fe/H] = $-2.0$ to 0.5. We find that our
data points somewhat deviate from W94 model. That is, our data are
somewhat bluer in $V-I$ than W94 model as a whole. The discrepancy
would be ascribed to the assumption of W94 model considering only a
single stellar population of old age ($>1.5$ Gyr) and of low mass
stars $<2.0 M_{\odot}$). If young massive stars exist in the galaxy,
the $V-I$ would be bluer than the W94 model but the $I-J$ would not be
considerably influenced, because the young massive stars radiate
mainly in bluer bands. We find that the data points do not
considerably shift into the direction of the absorption vector. The
exception is the bulge of dusty galaxy NGC 253 ($V-I=2.60$,
$I-J=1.69$), which locates in the direction of the absorption vector
(out of the diagram). In addition, the color of NGC 3079 bulge is
abnormal: $V-I=1.01$ and $I-J=2.56$. Both NGC 253 and NGC 3079 are
high-inclination (nearly edge-on) and the bulge parameters would be
seriously influenced by dust extinction. The bulge of NGC 6946
($V-I=1.96$, $I-J=0.93$) would be also reddened by dust. Comparing our
result with W94 models, we note that the most galaxies of
low-inclination except for NGC 6946 would not be seriously affected by
internal dust extinction.

\begin{figure}
  \begin{center} \FigureFile(80mm,80mm){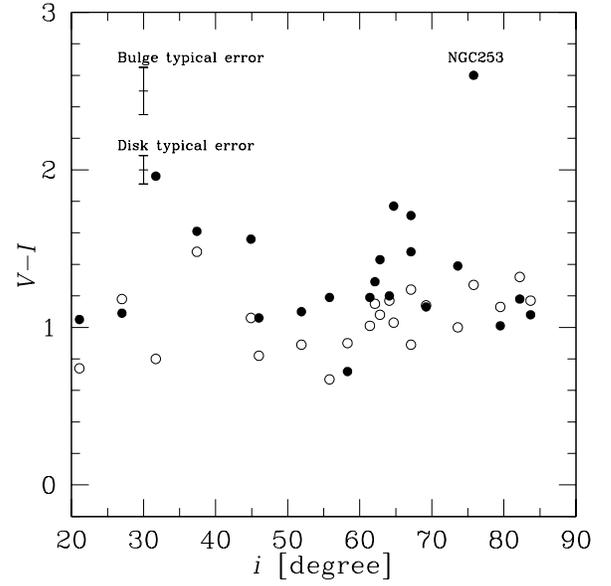} \end{center}
    \caption{The inclination $i$ versus $V-I$ diagram. The filled and
    open circles represent bulge and disk, respectively. }
    \label{fig.11}
\end{figure}

Figure \ref{fig.11} shows the diagram of $V-I$ color versus
inclination $i$, which is quoted from RC3. In general, since the
internal extinction is higher in the galactic center, the bulge
luminosity would be affected more strongly by the extinction than the
disk luminosity (e.g., \cite{peletier1999}). We find that the bulges
are generally somewhat redder than the disks.

\begin{figure}
  \begin{center} \FigureFile(80mm,80mm){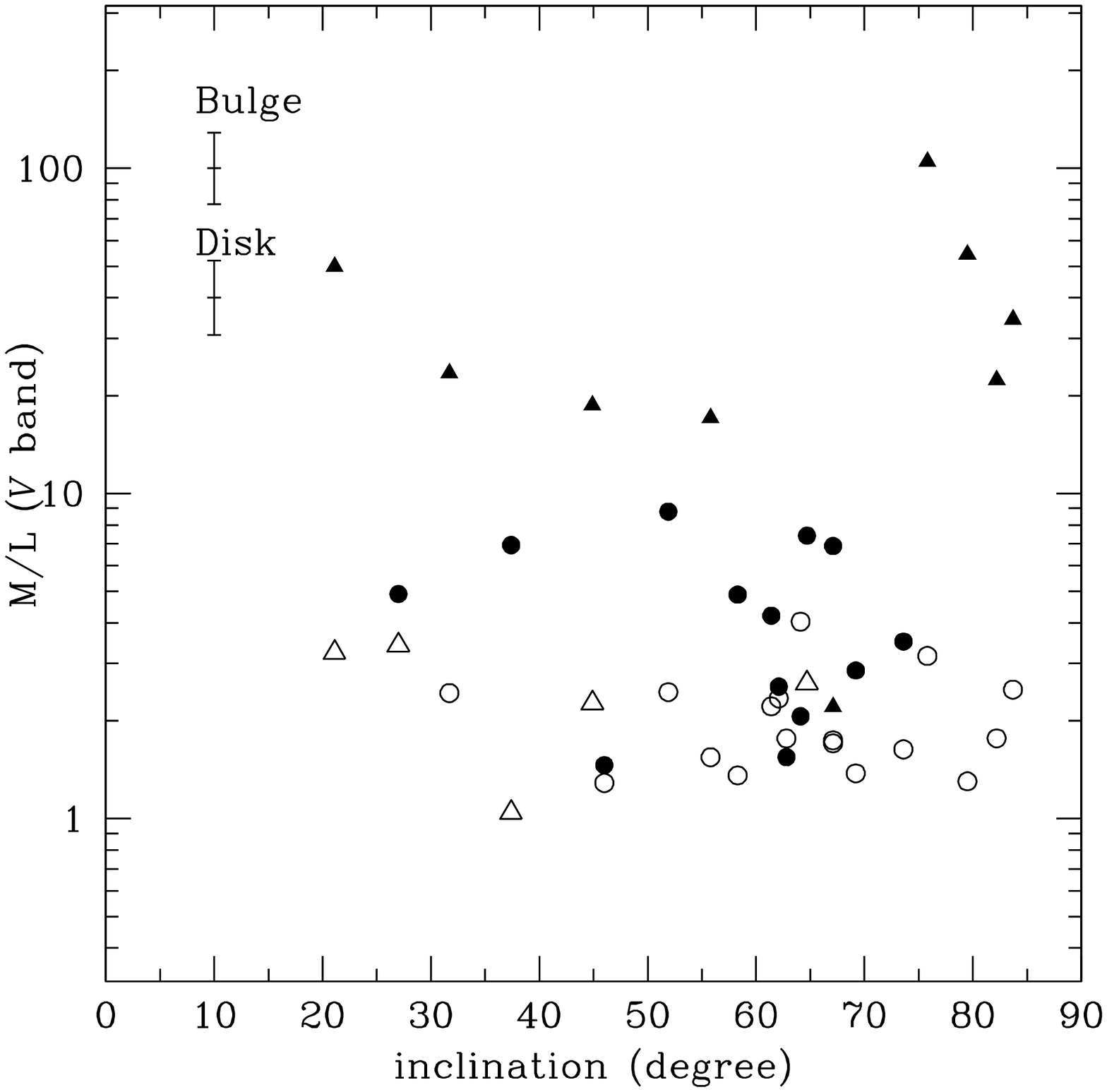} \end{center}
    \caption{The inclination $i$ versus $M/L_V$ diagram. The filled
    and open circles represent bulge and disk in the reliable range of
    rotation curve fitting ($\mu_{e,V}<20.5$, $\mu_{0,V}<20.5$),
    respectively. The filled and open triangles represent unreliable
    data of bulge and disk, respectively. } \label{fig.12}
\end{figure}

Figure \ref{fig.12} shows the diagram of $M/L_V$ ($M/L$ of $V$ band),
which would be sensitive to the dust extinction, versus inclination
$i$. The bulge $M/L_V$ for some galaxies are significantly higher than
normal values predicted with various stellar population synthesis
models: about 1 to 10. The high values (about 20 to 100) would be
caused by two reasons; One is the dust extinction, and the other is
the error of model fitting for rotation curve.  The former reason
would be valid for bulges of nearly edge-on ($i>75^{\circ}$) galaxies
(NGC 253, NGC 3079, NGC 4565 and NGC 5907), however the color is not
significantly red except for NGC 253.  The latter reason would be
suitable for most of those data, because the bulge effective magnitude
$\mu_{e,V}$ is significantly large (NGC 253, 2403, 3079, 3198, 4535,
4565, 5457, 5907 and 6946) and thus unreliable as mentioned in section
3.4. \citet{takamiya2000} has claimed that the high bulge $M/L$ for
low luminosity galaxies may be caused by the dark matter concentration
in the core region other than old stars or dust extinction. To focus
on stellar population, however, we use only reliable data. We consider
that the bulge $M/L$ or bulge $V-I$ outside the reliable range of
bulge $\mu_{e}$, disk $\mu_{0}$ or with $i>75^{\circ}$ would be
influenced by error of fitting or dust extinction, and thus are
excluded from the following discussion.

For the reliable sample, the average values of bulge $M/L$s for $V$,
$I$, $J$ bands are $4.5\pm2.3$ (13 sample), $2.7\pm1.7$ (14 sample)
and $1.0\pm0.2$ (5 sample), respectively. They do not include the
nearly edge-on ($i>75^{\circ}$) galaxies. The disk $M/L$s for $V$,
$I$, $J$ bands are $2.0\pm0.7$ (17 sample), $1.4\pm0.5$ (21 sample)
and $0.7\pm0.3$ (11 sample), respectively. The average of bulge $V-I$
is $1.33\pm0.30$ (18 sample) and that of disk $V-I$ is $1.05\pm0.20$
(22 sample). 

\section{Discussion}
\subsection{The Comparison with Galaxy Evolution Models}

\begin{figure*}
  \begin{center} \FigureFile(170mm,170mm){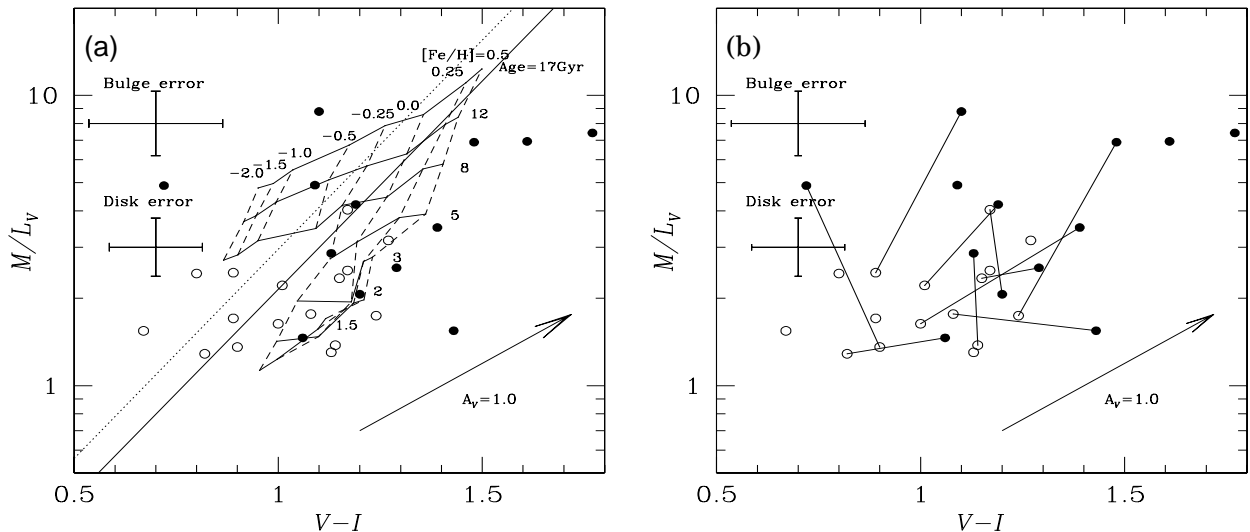} \end{center}
   \caption{The color-$M/L$ diagram in the $V$ band. The filled and open
   circles represent the bulges and the disks, respectively. Only
   reliable data are plotted. The error bars represent the typical
   errors for bulge and disk. (a) : Comparison with the models. The
   different models are as follows: GISSEL96 with Scalo IMF (solid
   line), with Salpeter IMF (dotted line). The models of W94 of the
   same age are connected by solid lines, while the models of the same
   metallicity are connected by dashed lines. (b) : Comparison the
   bulge with the counterpart disk for the same galaxy, by connecting
   with solid lines. Only reliable data for both the bulge and disk in
   the same galaxy are connected. } \label{fig.13}
\end{figure*}

\begin{figure*}
  \begin{center}
    \FigureFile(170mm,170mm){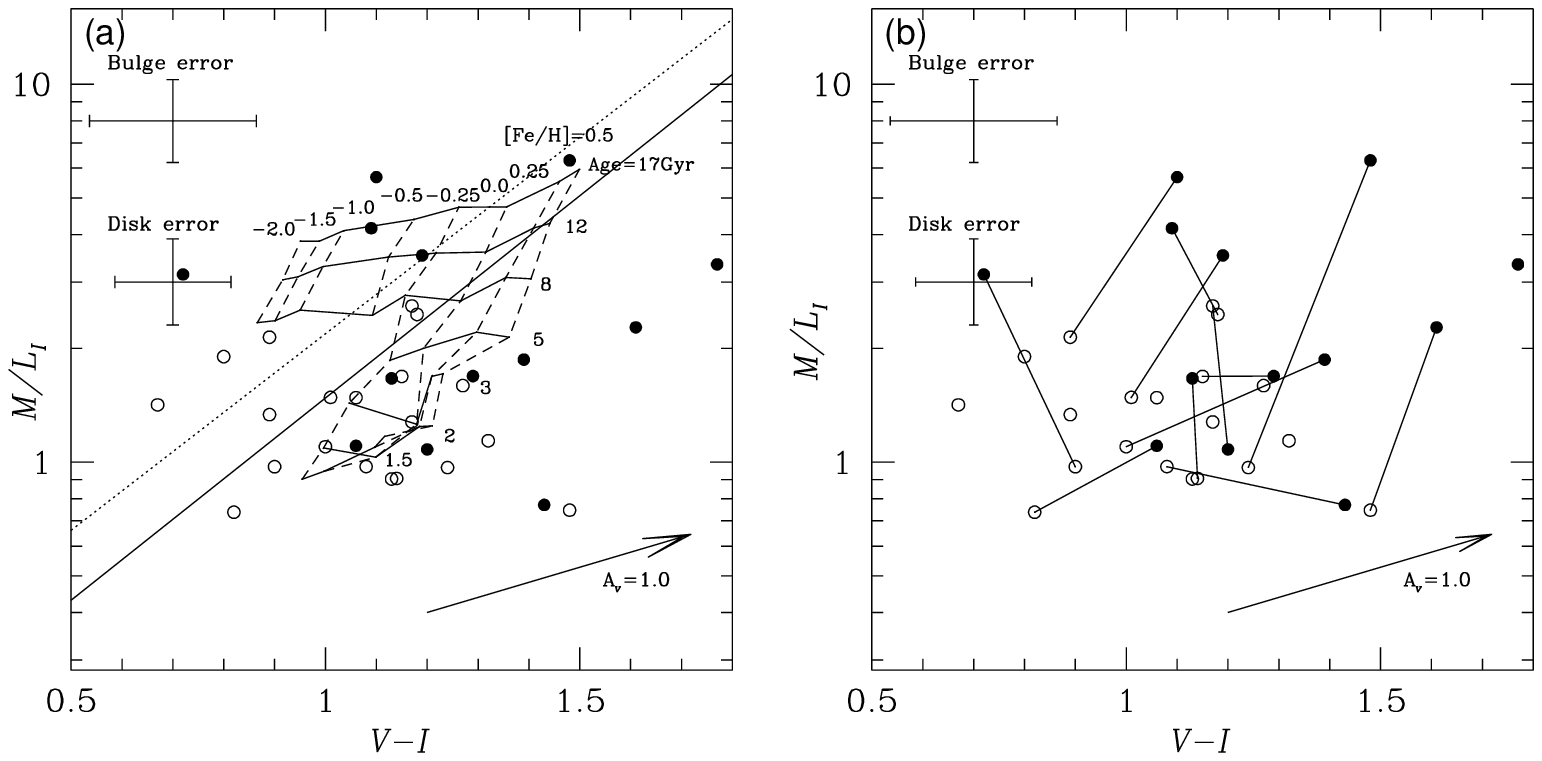}
  \end{center}
  \caption{The color-$M/L$ of $I$ band diagram. The symbols are the same
 as for figure \ref{fig.13}.} 
 \label{fig.14}
\end{figure*}

\begin{figure*}
  \begin{center}
    \FigureFile(170mm,170mm){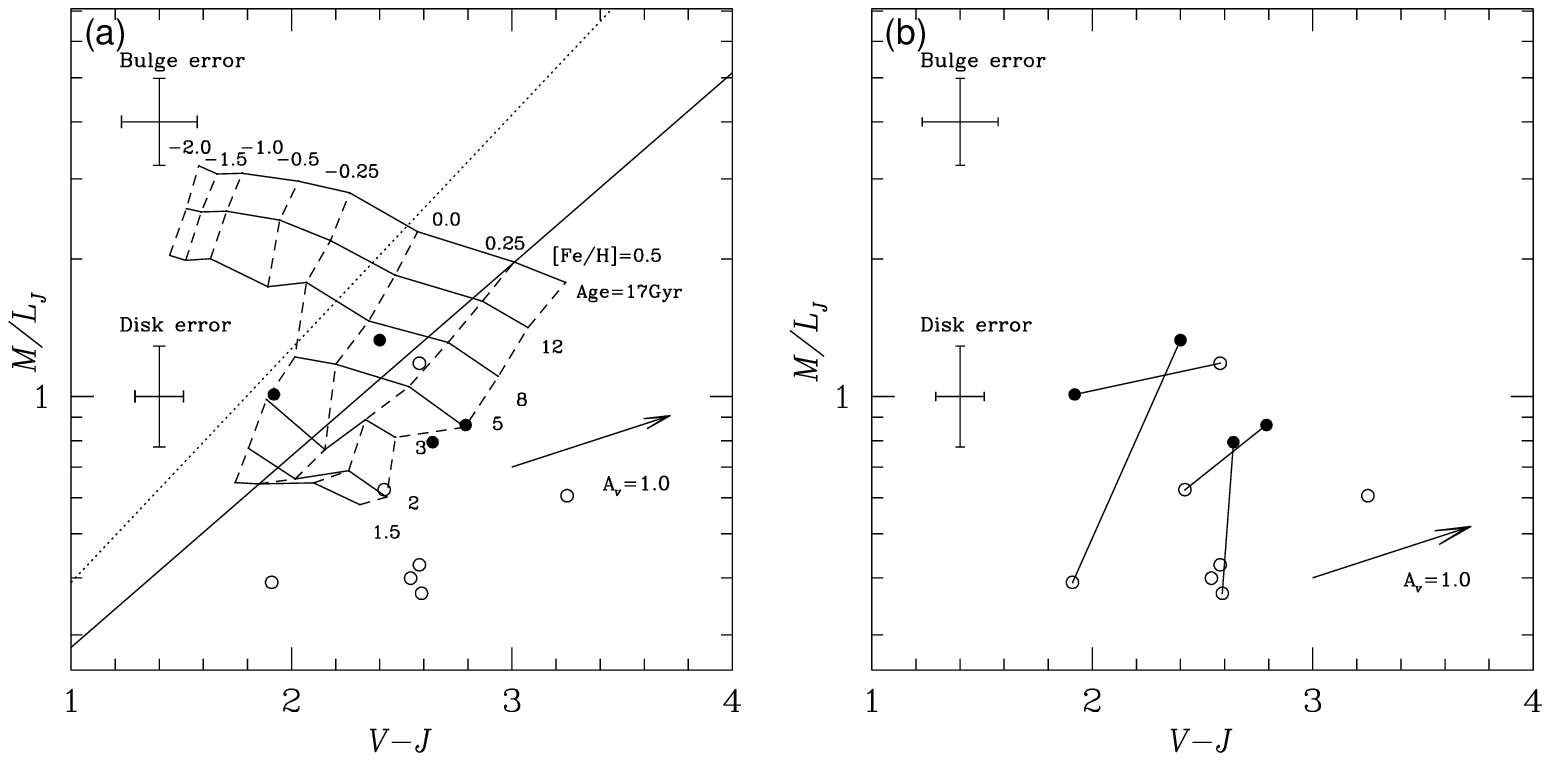}
  \end{center}
  \caption{The color-$M/L$ of $J$ band diagram. The symbols are the same
 as for figure \ref{fig.13}.} 
 \label{fig.15}
\end{figure*}

We compare the obtained colors and $M/L$s with galaxy evolution models
in figures \ref{fig.13}-\ref{fig.15}. Only reliable data as mentioned
in the previous section are plotted.  The mesh in the figure
represents W94 model of single starburst formation with Salpeter IMF,
the age = 1.5 to 17 Gyr, and [Fe/H] = $-2.0$ to 0.5. When we compare
the obtained colors and $M/L$s with those of W94 model, we estimate
the luminosity-weighted average age and metallicity of the system
represented by the single star formation. The two lines represent the
model sequences taken from BD01, which are based on an exponentially
declining SFR model of GISSEL96 (see \cite{bruzual1993};
\cite{leitherer1996}) with various time-scale $\tau$ and IMFs. Here we
show the two models with IMFs of \citet{salpeter1955} and
\citet{scalo1986} in the figures. The formation epoch and the
metallicity are 12 Gyr and Z=0.02, respectively, and $\tau$ varies
from 1 to infinity for both of two models. The upper-right parts of
the lines correspond to the models of small $\tau$, which consist of
only old population, whereas the lower-left parts of the lines
correspond to the models of large $\tau$, which include relatively
young population. The color and $M/L$ inferred from small $\tau$
models are consistent with those from W94 model at 12 Gyr with
[Fe/H]=0.0.

In general, the $M/L$ is significantly sensitive to massive young
stars in the system. BD01 has claimed that a large secondary starburst
(10 \% mass of the galaxy changes into stars) decreases the system
$M/L$ at a given color by up to a factor of 3. In addition, the
difference of IMF also considerably affects the zero-point of the
color-$M/L$ correlation, especially in near-infrared band, while the
choice of stellar population synthesis model does not notably change
the zero-point. The sensitivity of the zero-point of the color-$M/L$
correlation to the IMF is due entirely to differences in the number of
low-mass stars in each IMF (BD01). Since Scalo IMF produces less
low-mass stars than Salpeter IMF, the model with Scalo IMF yields
lower $M/L$ at a given color than the model with Salpeter IMF.

It is known that continuous star forming models also give the similar
result as the model of shallower slope IMF $+$ exponentially declining
SFR, however BD01 have not give the values of $M/L$s and colors using
continuous star forming models. Note that continuous star formation
models may be suitable as well as the models of shallower slope IMF
$+$ exponentially declining SFR.

We find in the (a) part of figures \ref{fig.13}-\ref{fig.15} that the
color and $M/L$s for both bulges and disks are generally in good
agreement with the models. In general bulges have higher $M/L$s and
redder color than disks in the figures. The (b) part of figures
\ref{fig.13}-\ref{fig.15} show the comparison of the bulge with the
counterpart disk for the same galaxy in the $M/L$-color diagrams. We
find that the bulges have generally higher $M/L$ and redder color than
the counterpart disk in the galaxy.  Although the trend would be
partially caused by the dust extinction, the direction of the
connected lines of the galaxies are somewhat different from the
extinction vector in many cases. We therefore conclude that bulges are
generally older than disks. However, some bulges have almost the same
or lower $M/L$s and colors as disks. These bulges should be as young
as disks.
 
The effects of age, metallicity, IMF and dust extinction almost
degenerate in the $V$ band (figure \ref{fig.13}). On the other hand,
the degeneracy of age, metallicity and dust extinction is resolved in
the $I$ band (figure \ref{fig.14}), and the difference of IMF is
resolved in the $J$ band (figure \ref{fig.15}), in which the effect of
dust extinction is smaller than in the $V$ band. Most of bulges and
disks agree with the model with the exponentially declining SFR.  In
addition, the Scalo or shallower slope IMF is appropriate in the $J$
band (figure \ref{fig.15}), though the data points are scarce. The
$M/L$ values of W94 model at the $J$ band are somewhat different from
our results. The discrepancy would be due to the IMF used in W94; the
Salpeter IMF tends to produce higher $M/L$ than Scalo or other
shallower slope IMFs at the $J$ band rather than at the $V$ and $I$
band. The figures \ref{fig.13}-\ref{fig.15} indicate that the model of
exponentially declining SFR $+$ shallower slope IMF (or continuous
star formation models) is preferable for both bulge and disk, rather
than the model including only old age stars and steeper slope IMF.  We
emphasize that the models of long term star-forming is preferable not
only for disks but also for most of bulges. Note that the same result
has been reported in BD01 for disks.

\begin{figure*}
  \begin{center} \FigureFile(170mm,170mm){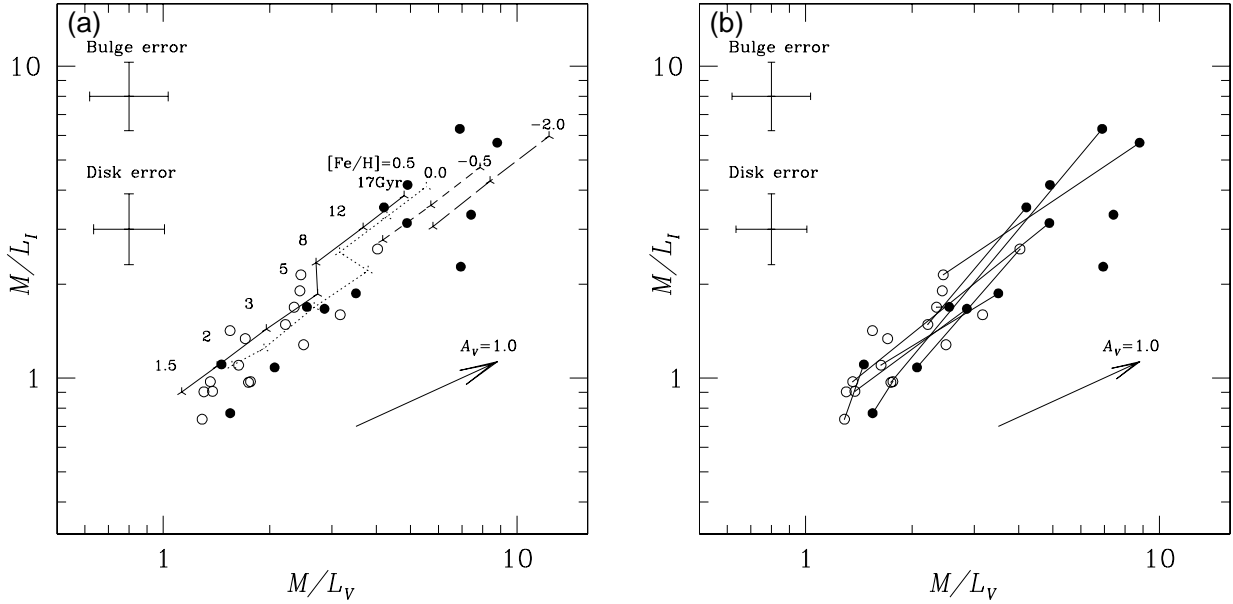} \end{center}
    \caption{The $M/L_V$-$M/L_I$ diagram. The filled and open circles
    represent the bulges and the disks, respectively. Only reliable
    data are plotted. (a) : Comparison with models. The models of W94
    of [Fe/H]=0.5, 0.0, -0.5 and -2.0 are shown. The dots represent
    ages of 1.5, 2, 3, 5, 8, 12 and 17 Gyr. (b) : Comparison the bulge
    with the counterpart disk for the same galaxy, by connecting with
    solid lines. Only reliable data for both the bulge and disk in the
    same galaxy are connected. } \label{fig.16}
\end{figure*}

\begin{figure*}
  \begin{center} \FigureFile(170mm,170mm){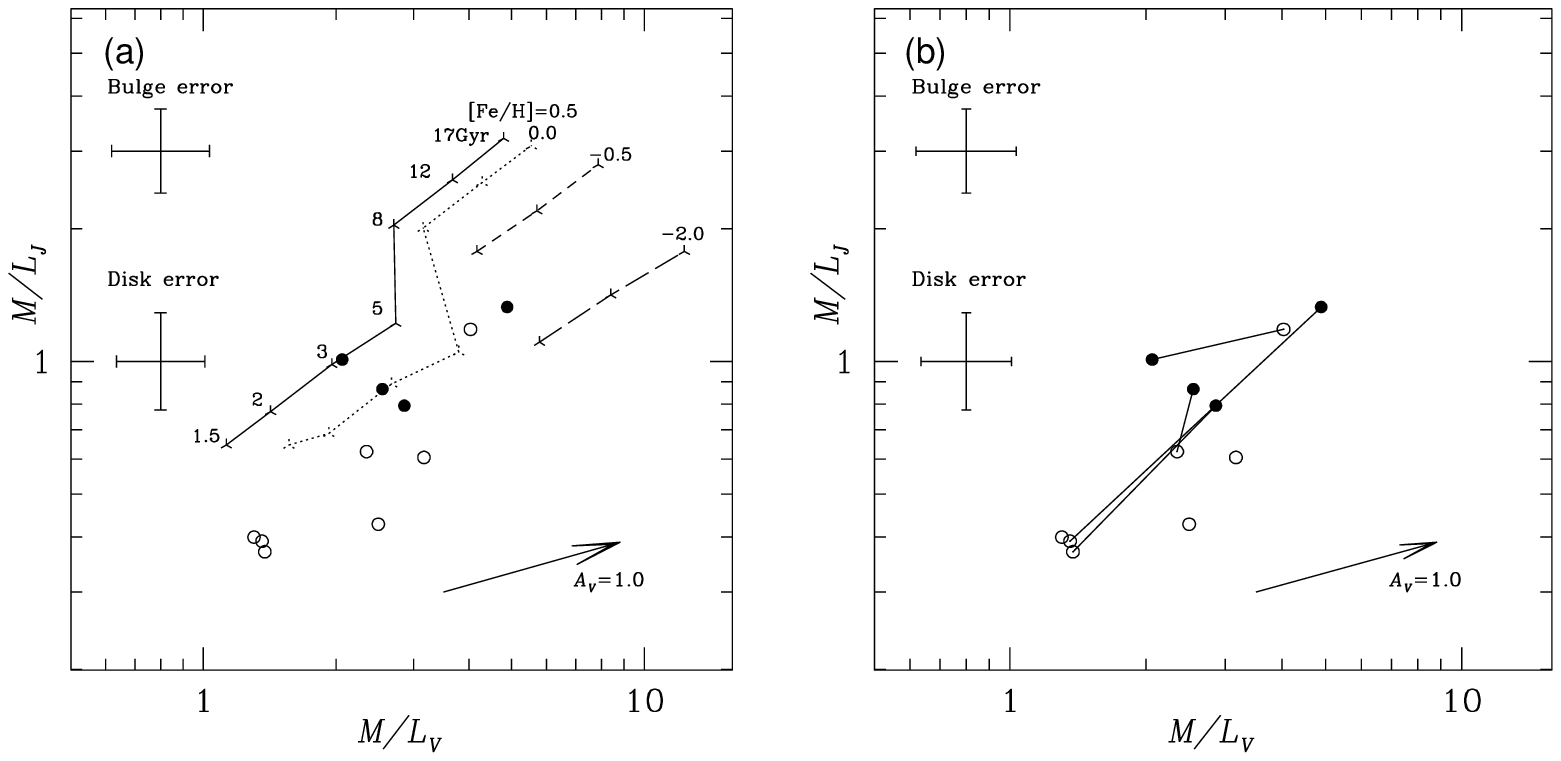} \end{center}
    \caption{The $M/L_V$-$M/L_J$ diagram. The symbols are the same
 as for figure \ref{fig.16}.}
    \label{fig.17}
\end{figure*}

To check the validity of the $M/L$s, we illustrate the $M/L_V$
vs. $M/L_I$ diagram and the $M/L_V$ vs. $M/L_J$ diagram in figure
\ref{fig.16} and \ref{fig.17}, respectively. The W94 models of single
starburst with Salpeter IMF, the age= 1.5 to 17 Gyr, and [Fe/H] =
$-2.0$ to 0.5 are plotted in the (a) part of figures \ref{fig.16} and
\ref{fig.17}. We also illustrate the comparison of the bulge the
counterpart disk in the same galaxy in the (b) part of figure
\ref{fig.16} and \ref{fig.17}. We find that the obtained $M/L$s for
both of bulges and disks agree with the model lines of the age
sequences, and the bulges are generally older than the counterpart
disks in the figure \ref{fig.16} and \ref{fig.17}. The ages of bulges
and disks are not uniform but widely distributed. It is clear that the
variety of $M/L$ values for both bulge and disk is caused mainly by
the age rather than by the metallicity. Moreover, the distribution of
bulge age is wider than that of disk; some bulges are older and the
others are as same as or younger than disks. The age effect would be
dominant rather than the metallicity effect and the dust extinction in
these figures, however the degeneracy is not entirely solved. At
least, these figures show that some bulges have low $M/L$s and
therefore these bulges are young even if the dust extinction or the
variety of metallicity also cause the wide distribution of bulge
$M/L$s.

Our result is contrary to the prediction of monolithic collapse model,
in which the bulges were formed from the primordial gas with single
starburst, thus must be uniformly old. Our result shows that the ages
of bulges are distributed widely, however they are generally older
than the disks. On the other hand, in the secular evolution scenario
the pseudo bulge is formed gradually and include young stars via the
secondary star formation in addition to the classical bulge component
of old age. Consequently our result suggests that the secular process
(formation and evolution) scenario is valid at least for some
galaxies.

\subsection{Correlation between Morphological Type and $M/L$}

\begin{figure}
  \begin{center} 
   \FigureFile(80mm,80mm){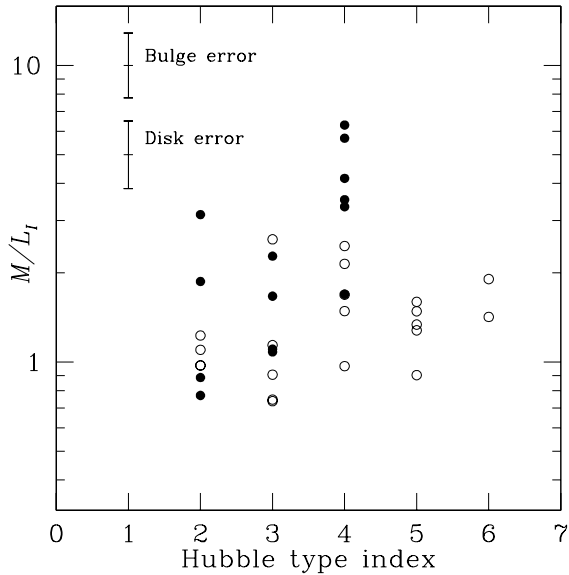} \end{center}
    \caption{The Hubble type index versus $M/L$ in the $I$ band. The
    filled and open circles denote the bulges and the disks,
    respectively. Only reliable data are plotted. }  \label{fig.18}
\end{figure}

\begin{figure}
  \begin{center}
    \FigureFile(80mm,80mm){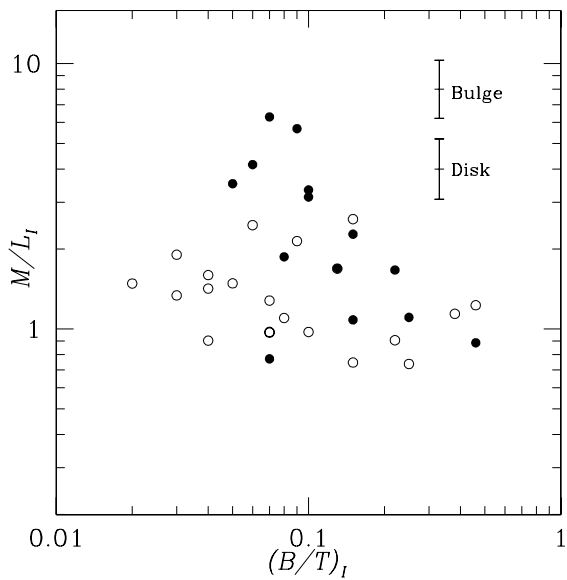}
  \end{center}
  \caption{The $B/T$ versus $M/L$ in the $I$ band. The symbols are the
 same as for figure \ref{fig.18}. Only reliable data are plotted.}
 \label{fig.19}
\end{figure}

To investigate the bulge evolution with morphology, we illustrate
diagrams of Hubble type index versus $M/L_I$ in figure \ref{fig.18}
and the bulge-to-total luminosity ($B/T$) ratios in the $I$ band
versus $M/L_I$ in figure \ref{fig.19}. We find in these figures that
the bulge $M/L$s increase with increasing Hubble type index or
decreasing $B/T$, while the disk $M/L$s do not correlate with the type
or $B/T$. Unfortunately the bulge $M/L$ of later-type (Sc or later) in
figure \ref{fig.18} is lack because the reliable data of bulge $M/L$
do not exist in the later-type. The trend do not change even if the
unreliable data are included, however. The trend is more obvious in
figure \ref{fig.19} ($B/T$ v.s. $M/L$ diagram) than figure
\ref{fig.18} (type index v.s. $M/L$ diagram). We consider that the
$B/T$ obtained directly from the decomposition is more accurate than
the type index classified by eye. The correlation suggests that bulges
of earlier-type spirals (Sab and Sb) are younger than those of
later-type spirals (Sbc or later), as \citet{kauffmann1996} has
predicted. We therefore consider that the secondary star formation
occurs in bulges of earlier-type spirals rather than those of
later-type spirals. Probably the bulge in the galaxy would be the sum
of the classical bulge component (formed from the rapid process at
high-z) and the pseudo bulge component (formed from the slow process
until today). The pseudo bulge component would be dominant in the
earlier-type spirals while the classical bulge component would be
dominant in the later-type spirals. However, since the number of $M/L$
sample is small and limited to the type of Sab-Sbc, further
observation is necessary to confirm the correlation.

\subsection{Correlation between Bulge Parameters and Disk Parameters}

\begin{figure}
  \begin{center} \FigureFile(80mm,80mm){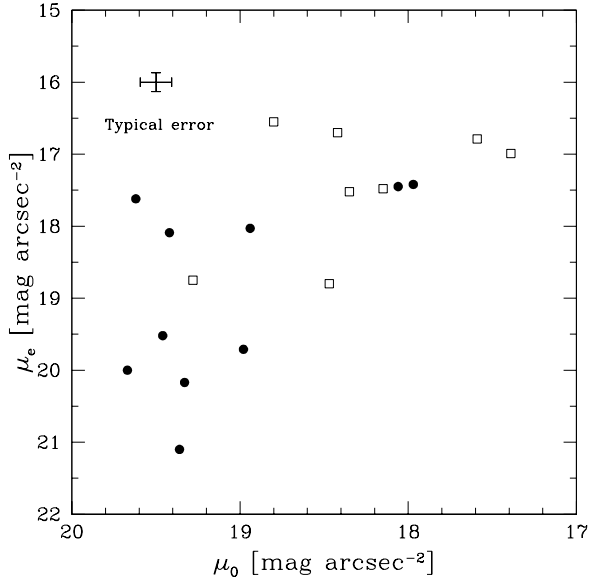} \end{center}
    \caption{The comparison of disk $\mu_0$ with bulge $\mu_e$ for $I$
    band. The squares and circles represent early ($T\le3$) and late
    ($T\ge4$) type galaxies, respectively.}  \label{fig.20}
\end{figure}

\begin{figure}
  \begin{center} \FigureFile(80mm,80mm){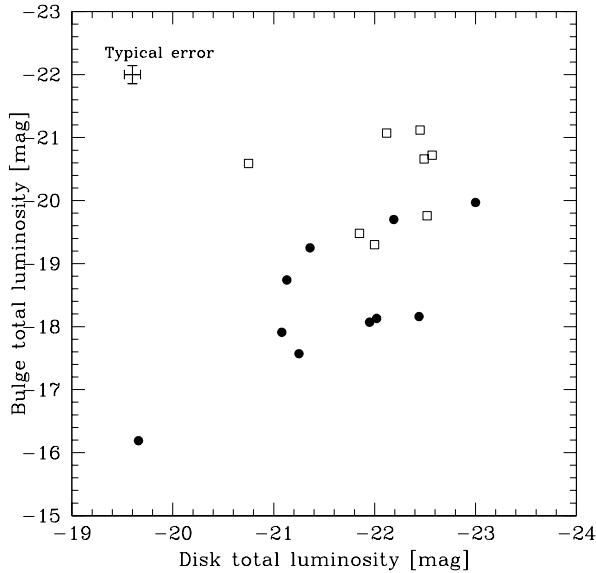} \end{center}
    \caption{The comparison of disk total luminosity with bulge total
    luminosity for $I$ band in absolute magnitude unit. The squares
    and circles represent early ($T\le3$) and late ($T\ge4$) type
    galaxies, respectively. The error bar does not include the error
    of estimating distance.}  \label{fig.21}
\end{figure}

Previous works (e.g., \cite{dejong1996b}, \cite{mollenhoff2001},
\cite{macarthur2003}, \cite{hunt2004}) reported that there are some
correlations between the structural parameters of bulge and disk, and
considered the correlation as the evidence of secular evolution of
bulge. We illustrate the disk central luminosity $\mu_0$ versus the
bulge effective luminosity $\mu_e$ in figure \ref{fig.20}. Assuming
that the bulges are formed separately from the disks as monolithic
collapse scenario predicts, no correlation between the bulges and the
disks should be found.  The bulge effective luminosity correlates with
the disk central luminosity in figure \ref{fig.20}, however. In
addition, most of earlier type (T=2, 3) galaxies locate in the
upper-right of the diagram, while later type galaxies (T=4, 5, 6)
galaxies locate in the lower-left. That is, the early type spirals
have generally brighter $\mu_e$ and $\mu_0$ than the late type
spirals. Figure \ref{fig.21} shows the disk total luminosity versus
the bulge total luminosity. The same tendency as for figure
\ref{fig.20} is seen in figure \ref{fig.21}. This result agrees with
\citet{dejong1996b}.

\begin{figure}
  \begin{center} \FigureFile(80mm,80mm){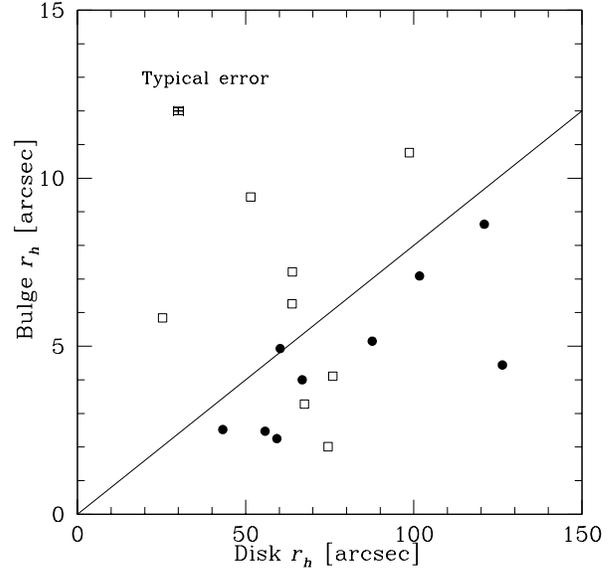} \end{center}
    \caption{The comparison of scale lengths between bulge and disk in
    $I$ band.  The open squares and filled circles represent early
    ($T\le3$) and late ($T\ge4$) type galaxies, respectively.  The
    solid line shows $r_{hb}=0.08r_{hd}$.}  \label{fig.22}
\end{figure}

We also confirm the correlation between scale lengths of bulge and
disk, which is reported in previous works. The bulge scale length
$r_{hb}$ is transformed from the bulge effective radius $r_e$ using
the equation $r_{hb}=r_{e}/(1.9986/\beta - 0.327)^{1/\beta}$ (e.g.,
\cite{moriondo1998a}, \cite{mollenhoff2001}).  Figure \ref{fig.22}
shows the comparison between the bulge scale length $r_{hb}$ and that
of disk $r_{hd}$ for the $I$ band.  Although the dispersion is large,
the $r_{hb}$ is roughly proportional to the $r_{hd}$ and
$r_{hb}/r_{hd}$ is independent of the Hubble type.  We obtain the
ratio $r_{hb}/r_{hd}=0.08\pm0.05$.  Our result agrees with the
previous works.  \citet{courteau1996} reported that it is
$0.09\pm0.04$ for the galaxies of \citet{dejong1996b}, and
\citet{macarthur2003} reported that it is $0.13\pm0.06$ for 121
late-type galaxies. \citet{courteau1996} claimed that the constant
$r_{hb}/r_{hd}$ indicates the signature of secular formation because a
numerical simulation of disk galaxies evolve toward a double
exponential profile with a typical ratio between bulge and disk scale
length near 0.1. However, the bulge scale length $r_{hb}$ depends on
the value of $\beta$. Being adopted $\beta=1/4$ (de Vaucouleurs' law)
instead of $\beta=1$, the ratio of bulge/disk radii is significantly
changed, as \citet{moriondo1998a}, \citet{graham2001} and
\citet{hunt2004} have pointed out. \citet{hunt2004} reported that the
tight correlation is found for $\beta=1$ bulges and disks, but it
disappeared for $\beta=1/3$ and $\beta=1/4$ bulges and disks. The
adopted $\beta$ is nearly 1 for many galaxies in this paper as well as
\citet{dejong1996b}, hence the similar results as \citet{courteau1996}
would be obtained.

\begin{figure}
  \begin{center} \FigureFile(80mm,80mm){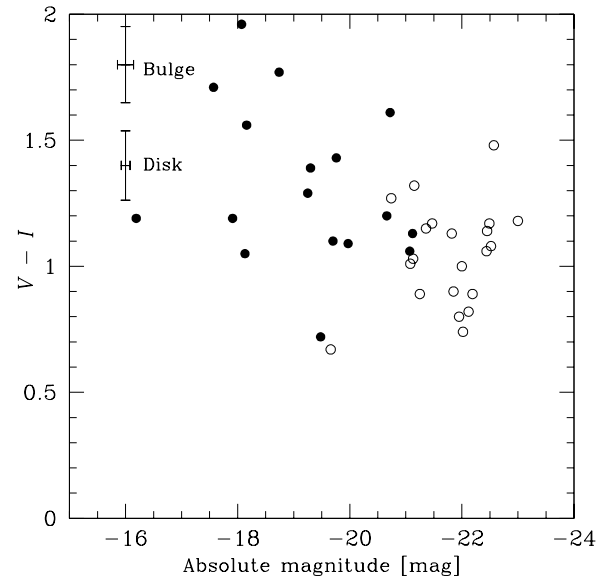} \end{center}
   \caption{The color-magnitude diagram. The $V-I$ color and $I$ band
   absolute magnitude of bulge and disk are shown. The filled and open
   circles represent the bulges and the disks, respectively. The error
   bar does not include the error of estimating distance.}
   \label{fig.23}
\end{figure}

Moreover, we investigate the color-magnitude relation in figure
\ref{fig.23}. If a tight correlation is found, the system should have
an age and a metallicity similar to elliptical galaxies. There is no
tight correlation in the figure for both bulges and disks. This
implies that the age of bulge is not uniform. Although the dispersion
is large, the brighter bulges tend to have slightly bluer than the
fainter bulges, i.e., the brighter bulges would include young
stars. The result also implies the secondary star formation in the
bulges of earlier-type spirals.

\subsection{Comparison with The Previous Researches}

\begin{figure*}
  \begin{center} \FigureFile(170mm,170mm){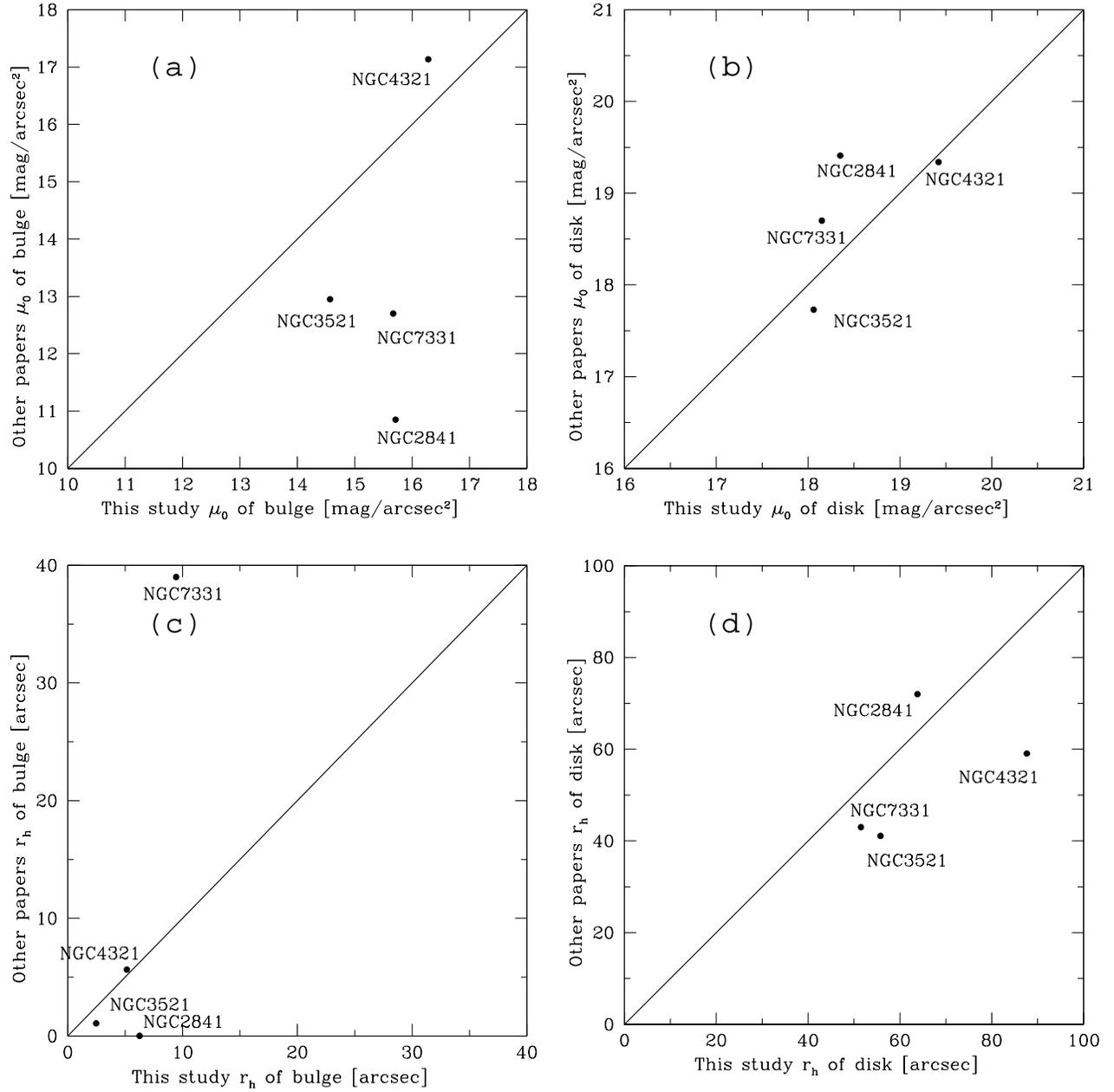} \end{center}
   \caption{The comparison of surface brightness parameters in this
   study with the previous papers. (a) central magnitude of bulge. (b)
   scale length of bulge. (c) central magnitude of disk. (d) scale
   length of disk. The solid lines in each panel show the equality.}
   \label{fig.24}
\end{figure*}

The surface brightness for some galaxies have been decomposed into
 bulge and disk in previous works. We compare the surface brightness
 parameters of the present study with those of previous works.  NGC
 2841 and NGC 3521 are decomposed by \citet{mollenhoff2004}, NGC 4321
 is decomposed by \citet{dejong1996a}, and NGC 7331 is decomposed by
 \citet{bottema1999} at the $I$ band. Note that various units of
 parameters for bulge are used in these papers; $\mu_e$ (effective
 magnitude) or $\mu_0$ (central magnitude), $r_e$ (effective radius)
 or $r_h$ (scale length), and kpc unit or arcsec unit. To avoid
 confusion, the scale length $r_{h}$ in arcsec unit and the central
 magnitude $\mu_{0}$ for bulge are used in this section instead of the
 effective radius $r_e$ and effective magnitude $\mu_e$. Figure
 \ref{fig.24} shows the comparison for the parameters at the $I$ band.

The disk parameters of the present paper are roughly consistent with
those of previous works in (b) and (d) of figure \ref{fig.24}.
However, the bulge parameters are inconsistent except for NGC 4321.
We consider that the adopted bulge index $\beta=1/n$ strongly affects
the values of structural parameters of bulge and therefore causes the
disagreement. In fact, NGC 4321 bulge is fitted with exponential
($\beta=1$) model in both \citet{dejong1996a} and this study, and the
values of parameters are roughly consistent. The $\beta$ is 0.25 for
the bulges of NGC 2841 and NGC 7331, and is 0.74 for NGC 3521 bulge in
previous works, while the $\beta$ is 1.0 for the bulges of NGC 2841
and NGC 7331 and is 0.67 for NGC 3521 bulge in this study.
\citet{baggett1998} also present the bulge/disk parameters for 17
galaxies in Table 4 of this paper. We have also compared with their
results, but the results are generally inconsistent. We consider the
inconsistent is due to the adopted $\beta$ value; \citet{baggett1998}
have adopted the $\beta=1/4$ for all galaxies while we fit mainly with
$\beta=1$.

The disk $M/L$s in the present study ($M/L_V=2.0\pm0.7$,
$M/L_I=1.4\pm0.5$, $M/L_J=0.7\pm0.3$) are somewhat lower than the
values of previous researches (e.g., \cite{moriondo1998b};
\cite{palunas2000}). The lower $M/L$s of this paper would be caused by
the following two reasons. The difference of assumed distances changes
$M/L$ because it is inversely proportional to the distance. In this
paper, the distances for 13 galaxies in table 2 are derived from the
Cepheid observation. Those for other 15 galaxies are cited from
\citet{tully1988} with $H_0 = 75$ ${\rm km~s^{-1}~Mpc^{-1}}$. On the
other hand, \citet{moriondo1998b} used the distances derived with $H_0
= 50$ ${\rm km~s^{-1}~Mpc^{-1}}$. If we adopt $H_0=72$ ${\rm
km~s^{-1}~Mpc^{-1}}$ (\cite{freedman2001}), the average of their disk
$M/L_J$ ratios is about 1.1 (originally 1.6). Thus the discrepancy is
reduced, however the disk $M/L_J = 0.7\pm0.3$ in the present study is
still lower than \citet{moriondo1998b}. The other reason of
discrepancy would be the correction of systematic error of the maximum
disk solution discussed in section 3.4. The corrected value of disk
$M/L$ is reduced by a factor of 0.9-1.0 compared with the original
value.

The bulge $M/L$s in the present study are $4.5\pm2.3$, $2.7\pm1.7$ and
$1.0\pm0.2$ in $V$, $I$ and $J$ band, respectively.  \citet{kent1986}
reported that the average bulge $M/L$ in $r$ band (an intermediate
band between $V$ and $I$) is 4.6 using optical emission line rotation
curves and $H_0=50$, which is 3.2 using $H_0=72$. In addition, using
H\emissiontype{I} rotation curves, \citet{kent1987} reported that it
is 2.9 ($H_0=72$).  Although the rotation curves used in
\citet{kent1986} were not the CO observation, these bulge $M/L$s agree
roughly with our result.  Moreover \citet{moriondo1998b} reported that
the average bulge $M/L$ in $J$ band is 1.1 ($H_0=50$), which is 0.76
using $H_0=72$. It is also roughly consistent with the bulge $M/L_J =
1.0\pm0.2$ in this study within 1 sigma. They claimed that the bulges
of early-type (Sa-Sb) are younger than disks, however the H${\alpha}$
rotation curves used in their studies are softened near the galactic
center. The CO rotation curves give more rapidly rotation curves than
the H$\alpha$ rotation curves near the galactic center, hence we tend
to obtain the higher bulge $M/L$ than the previous studies. The other
reason would be the correction for the systematic error of $M/L$; the
corrected value of bulge $M/L$ is about factor of 1.0-1.3 higher than
that of original value. If the correction is not applied, the bulge
$M/L$ is this paper is reduced by a factor of 0.7-1.0, and the
discrepancy between our result and \citet{moriondo1998b} would be 
reduced.

Using color-color diagram, \citet{peletier1999} reported that the
bulge colors of early-type (S0-Sb) are uniformly red and thus the ages
are as old as most of elliptical galaxies in cluster. However, W94
claimed that the colors are not sensitive to age except for extremely
young case ($<2$ Gyr) and are affected by the age-metallicity
degeneracy. Our bulge $V-I$ is $1.33\pm0.30$, which is somewhat redder
than disk $V-I=1.05\pm0.20$. Therefore our result is not inconsistent
with the report of \citet{peletier1999} for the bulge color. The $M/L$
has the advantage of being more sensitive to age than colors. Hence
the color-$M/L$ diagram reveals the spread of age for bulges rather
than the color-color diagram. Our result shows that the bulge age is
not uniform and that the bulges of earlier-type galaxies are younger
than those of later-type galaxies and nearly as same age as the disks.
However, the type of our bulge $M/L$ data is only from Sab to Sbc. Further
observation for S0-Sa and Sc-Sd galaxies are needed to confirm the
formation of bulges.

\section{Conclusion}

Using optical and near-infrared ($V$, $I$ and $J$) band images, we
decomposed the surface brightness of galaxies into bulge and disk
components. We obtained the $M/L$s for bulges and disks with CO
rotation curves for the inner part of galaxies. We compared the colors
and the $M/L$s with those of galaxy formation models. We found the
correlation between the colors and the $M/L$s for bulges and disks,
and that the correlation between $B/T$ and the $M/L$. Our results are
generally consistent with galaxy formation models of slowly
star-forming system; with an exponentially declining SFR and shallow
slope IMF (ex. Scalo IMF). The exponentially declining SFR taken from
BD01 is suitable not only for disks but also for bulges. Moreover we
found that the bulge $M/L$s of earlier-type spirals are lower than
those of later-type spirals. These facts suggest that the
luminosity-weighted average ages of the bulges in earlier-type
galaxies are younger than those of later-type galaxies. Moreover, we
confirmed the correlations between bulge parameters and disk
parameters as previous authors reported. In addition there is no tight
correlation between the colors and magnitudes for both bulges and
disks. These results suggest that the bulges of earlier-type galaxies
would not be the classical bulge but be the pseudo bulge as the
secular formation scenarios predict.  \\

We would like to thank an anonymous referee for careful reading of the
manuscript and valuable comment on this work. We wish to thank
Shin-ichi Ichikawa, Fumiaki Nakata and Yoshihiko Yamada of Astronomy
Data Center of National Astronomical Observatory of Japan for useful
comment. We also wish to thank Yumiko Tanaka of Kiso Observatory for
obtaining KONIC data.  This work is based on data collected at Kiso
observatory (University of Tokyo) and obtained from the SMOKA, which
is operated by the Astronomy Data Center, National Astronomical
Observatory of Japan. Data analysis was in part carried out on "sb"
computer system operated by the Astronomy Data Center of the National
Astronomical Observatory of Japan. This work has been supported in
part by the Grant-in-Aid for Scientific Research 11554005 and 14340059
of the Ministry of Education, Science, Culture, and Sports in Japan.

\end{document}